%% file: auctions.tex
\documentclass[sigplan,10pt,dvipsnames,table,xcdraw]{acmart}

\usepackage[utf8]{inputenc}
\usepackage{listings}
\usepackage{xcolor}
\usepackage{subcaption}
\usepackage{soul}
\usepackage{times}
\usepackage{array,multirow}
\usepackage{xspace}
\usepackage{balance}
\usepackage{paralist}
\usepackage{amsfonts}
\usepackage{amsthm,amsmath}
\usepackage{cleveref}
\usepackage{graphicx}
\usepackage{pifont}
\usepackage{array}
\usepackage{booktabs}
\usepackage{tikz}
\usepackage{bm}
\usepackage{enumitem}
\usepackage{todonotes}
\usepackage{enumitem}
\definecolor{LRed}{rgb}{1,.8,.8}
\definecolor{MRed}{rgb}{1,.6,.6}
\definecolor{HRed}{rgb}{1,.2,.2}

\usepackage[font=footnotesize,labelfont=bf]{caption}
\input{macros.tex}

\copyrightyear{2020}
\acmYear{2020}
\setcopyright{acmlicensed}\acmConference[Middleware '20]{21st International Middleware Conference}{December 7--11, 2020}{Delft, Netherlands}
\acmBooktitle{21st International Middleware Conference (Middleware '20), December 7--11, 2020, Delft, Netherlands}
\acmPrice{15.00}
\acmDOI{10.1145/3423211.3425669}
\acmISBN{978-1-4503-8153-6/20/12}

\ccsdesc[500]{Security and privacy~Trusted computing}
\ccsdesc[500]{Security and privacy~Distributed systems security}

\keywords{blockchains, security and privacy, trusted computing, distributed systems}

\begin{document}

\title{\sysname: Privacy-preserving, Auditable, Scalable \& Trustworthy Auctions for Multiple Items}

\author{Michał Król}
\affiliation{%
  \institution{City, University of London}
  \country{United Kingdom}}
\email{michal.krol@city.ac.uk}

\author{Alberto Sonnino}
\affiliation{%
  \institution{University College London}
  \country{United Kingdom}}
\email{alberto.sonnino@ucl.ac.uk}

\author{Argyrios Tasiopoulos}
\affiliation{%
  \institution{University College London}
  \country{United Kingdom}}
\email{argyrios.tasiopoulos@ucl.ac.uk}

\author{Ioannis Psaras}
\affiliation{%
  \institution{University College London}
  \country{United Kingdom}}
\email{i.psaras@ucl.ac.uk}

\author{Etienne Rivière}
\affiliation{%
  \institution{ICTEAM, UCLouvain}
  \country{Belgium}}
\email{etienne.riviere@uclouvain.be}

\begin{abstract}
	Decentralised cloud computing platforms enable individuals to offer and rent resources in a peer-to-peer fashion.
	They must assign resources from multiple sellers to multiple buyers and derive prices that match the interests and capacities of both parties.
	The assignment process must be decentralised, fair and transparent, but also protect the privacy of buyers.

	We present \sysname, a decentralised platform enabling trustworthy assignments of items and prices between a large number of sellers and bidders, through the support of multi-item auctions.
	\sysname uses threshold blind signatures and commitment schemes to provide strong privacy guarantees while making bidders accountable.
	It leverages the Ethereum blockchain for auditability, combining efficient off-chain computations with novel, on-chain proofs of misbehaviour.
	Our evaluation of \sysname using Filecoin workloads show its ability to efficiently produce trustworthy assignments between thousands of buyers and sellers.

\end{abstract}

\maketitle

\input{sections/introduction.tex}
\input{sections/goals.tex}

\input{sections/overview.tex}
\input{sections/background.tex}
\input{sections/preparation.tex}
\input{sections/execution.tex}

\input{sections/security.tex}

\input{sections/evaluation.tex}

\input{sections/related.tex}

\input{sections/limitations.tex}
\input{sections/conclusion.tex}

\section*{Acknowledgments}
This work is partially supported by the Belgian FNRS project DAPOCA (33694591), EPSRC INSP Early Career Fellowship under grant agreement number EP/M003787/1 and Facebook Calibra.

\appendix
\input{appendices/blind_bls.tex}

\bibliography{biblio}
\bibliographystyle{ACM-Reference-Format}

\end{document}

%% file: macros.tex
\newboolean{showcomments}
\setboolean{showcomments}{true}
\ifthenelse{\boolean{showcomments}}
{ \newcommand{\mynote}[3]{
   \fbox{\bfseries\sffamily\scriptsize#1}
   {\small$\blacktriangleright$\textsf{\emph{\color{#3}{#2}}}$\blacktriangleleft$}}}
{ \newcommand{\mynote}[3]{}}

\newcommand{\michal}[1]{\mynote{Michał}{#1}{blue}}

\newcommand\sysname{PASTRAMI\xspace}

\newcommand\ethereum{Ethereum\xspace}

\newcommand\algorithm[1]{\textsf{#1}}
\newcommand\hashtopoint{H^*\xspace}

\newcommand\bidder{bidder\xspace}
\newcommand\bidders{bidders\xspace}

\newcommand\Bidders{Bidders\xspace}
\newcommand\authority{authority\xspace}
\newcommand\authorities{authorities\xspace}

\newcommand\seller{seller\xspace}
\newcommand\sellers{sellers\xspace}

\newcommand{\etal}{\textit{et al.}\@\xspace}
\newcommand{\eg}{\textit{e.g.,}\@\xspace}
\newcommand{\ie}{\textit{i.e.,}\@\xspace}

\newcommand{\cmark}{\ding{51}}%
\newcommand{\xmark}{\ding{55}}

\def\first{({\it i})\xspace}
\def\second{({\it ii})\xspace}
\def\third{({\it iii})\xspace}

\newcommand\para[1]{\vspace{0.05in} \noindent \textbf{#1.}}
\newcommand\definition[2]{\ding{118}\xspace \textsf{#1}\xspace$\bm{\rightarrow}$\xspace(#2):\xspace}

\definecolor{verylightgray}{gray}{0.9}

\newcolumntype{L}{l<{\hspace{0.8cm}}}
\newcolumntype{C}{c}

\DeclareRobustCommand\pie[1]{
\tikz[every node/.style={inner sep=0,outer sep=0, scale=1.5}]{
\node[minimum size=1.5ex] at (0,-1.5ex) {}; 
 \draw[fill=white] (0,-1.5ex) circle (0.75ex); \draw[fill=black] (0.75ex,-1.5ex) arc (0:#1:0.75ex); 
}
}
\def\L{\pie{0}} 
\def\M{\pie{-180}} 
\def\H{\pie{360}} 

%% file: sections/introduction.tex


\section{Introduction}
\label{sec:introduction}


Sharing economy systems allow individuals to rent or ``share'' their resources to other individuals in a peer-to-peer fashion.
Multiple platforms already implement this concept towards a decentralised form of cloud computing.
They allow renting the machines of other users to run large-scale computations (e.g. Golem~\cite{golem}, iExec~\cite{iexec}, Pando~\cite{lavoie2018pando} or SONM~\cite{sonm}) or for storing data (e.g. Storj~\cite{storj} or Filecoin~\cite{filecoin}).
These platforms often build on blockchains as a global source of trust and as an enabler of decentralised payments.
For instance, the decentralised file storage platform Filecoin~\cite{filecoin} encrypts and scatters end-users' files fragments across multiple storage machines and rewards them with periodic token payments. 


One of the main challenges faced by sharing economy platforms is the difficulty of setting up fair allocations and prices between large numbers of sellers and buyers of goods or services (\emph{items} in the rest of this paper).
They need to set up decentralised mechanisms to \emph{globally} decide on the assignment of items from sellers to buyers, and the prices that apply to the transactions.
The interests of buyers and sellers are, indeed, in conflict: A buyer wishes to buy an adequate item at a minimal price, while a seller wants to sell its item(s) at the highest possible price. Direct negotiations between the two parties, as used currently in Filecoin~\cite{filecoin}, have poor scalability: Each buyer may have to individually contact and negotiate with an increasing number of storage nodes as competition for items increases. Furthermore, direct negotiation does not provide transparency and cannot guarantee any fairness: there are no guarantees that a seller or buyer settles the most interesting of all possible deals.

\begin{figure}[b!]
\includegraphics[width=\linewidth]{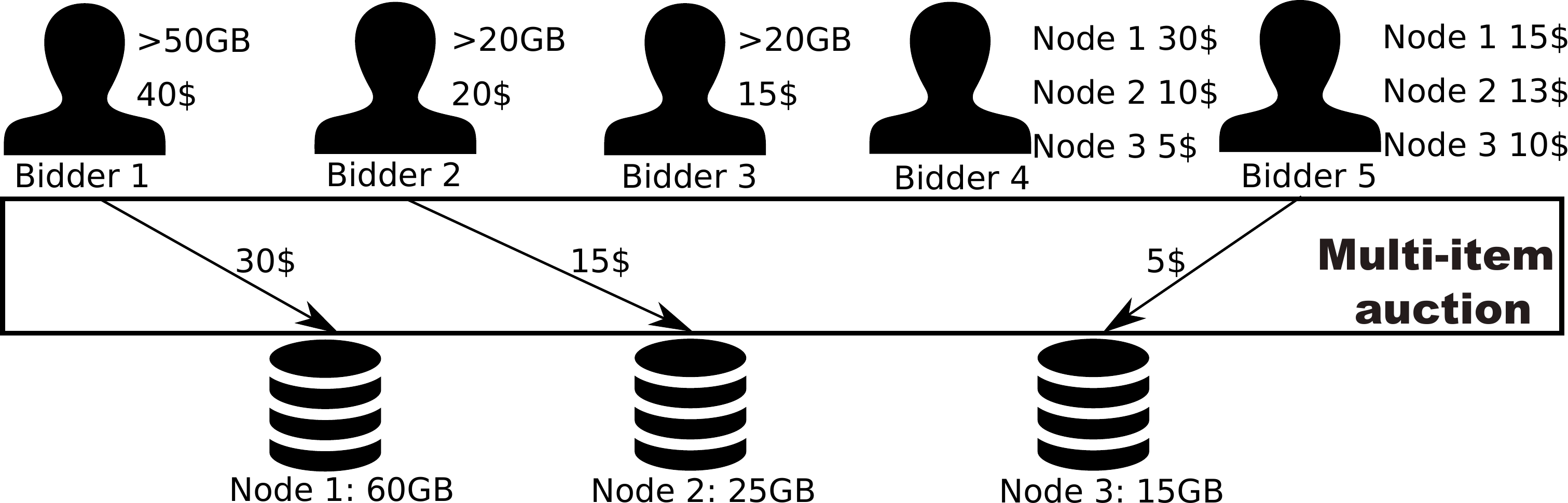}
\caption{Example of general (bidders 1--3) and specific (bidders 4 \& 5) bids for multiple items in the context of a decentralised storage cloud, and of the assignment of items and prices decided by a multi-item auction algorithm.}
\label{fig:intro}
\end{figure}

Multi-item auctions are a sound basis for the assignment of items and prices in decentralised exchange platforms.
They allow buyers, or \emph{bidders}, to announce the price they are willing to pay for an item (their \emph{valuation}) across many items from multiple sellers.
Valuations may be for a \emph{specific} item or for any one of the items that satisfy a number of constraints.
We denote the latter case as a \emph{general bid}.
We present an example in Figure~\ref{fig:intro}, inspired by Filecoin: Bidder~1 is willing to pay up to 40~\$ for a node with \emph{at least} 50~GB of storage space, while bidder~5 specifically bids for the three available storage nodes.
Based on these specific and general bids, a multi-item auction algorithm can derive an assignment (e.g., bidder~2 is assigned node~2) and prices for the transactions (e.g., bidder~1 has to pay 30~\$, less than the announced 40~\$).
The assignment of items and prices depend on a global measure of goodness that is specific to the multi-item auction algorithm used.
In this example, resources are assigned to bidders who ``value them'' the most while deciding on a price following rules of supply and demand.

The pricing mechanism is a critical component of an exchange platform and all participants should be convinced of its \emph{correctness}. A single malicious auction operator could easily influence the assignment to maximise its own gain. At the same time, verification should not require all participants to reproduce the entire computation.

Transparency and verification come, however, in apparent tension with the need to preserve the \emph{privacy} of auctions' participants.
First, \emph{sealed-bid auctions} require to keep bids private until all of them are submitted.
This prevents late bidders from taking advantage from the already revealed information~\cite{harkavy1998electronic}.
Second, it is important to keep bidders' identity private. For instance, revealing that a company recently rented significant amounts of storage may expose operation details of this company to its competitors.

Finally, it is necessary to compute auction results in an \emph{efficient} and \emph{scalable} manner. Auction algorithms experience exponential increase in execution time and require storing significant volumes of data with increasing size of the input parameters. In sharing economy systems, potentially thousands of buyers must be assigned to sellers on a regular basis.

Decentralised auctions have been realised previously using Secure Multi-Party Computations (MPC)~\cite{andrychowicz2014secure, bogetoft2009secure}.
MPC-based auctions provide strong privacy and correctness guarantees, but introduce significant computational overhead and can be executed only between a limited number of parties.
It means that the computations have to be performed by a small number of actors that must be trusted by all of the auction participants, something that is difficult to realise in practice. 

Blockchain platforms with support for executable smart contracts are an appealing solution for building auditable auctions, thanks to the immutability of their decentralised ledger~\cite{blass2018strain,galal2018verifiable,galal2018succinctly}.
However, direct auction execution using a smart contract impairs privacy and is at odds with scalability: The costly assignment calculation has to be repeated by all miners, leading to large overheads.
General platforms for verifiable computations such as Truebit~\cite{teutsch2016truebit} or Arbitrum~\cite{kalodner2018arbitrum} reduce the overhead, but require multiple verifiers to repeat full programs and do not offer sufficient privacy protection. 
While many specific auctions systems try to increase bidders' privacy or offload heavy computations off-chain, they require a trusted third party~\cite{blass2018strain}, secure hardware~\cite{yuan2018shadoweth} or involve significant computational overhead~\cite{galal2018verifiable}.

\para{Contributions}
We present the design and evaluation of \sysname, a framework for \emph{private}, \emph{secure} and \emph{efficient} multi-item auctions.
\sysname leverages the Vicrey-Dutch algorithm (VDA)~\cite{mishra2004multi,mishra2009multi} to determine an assignment between bidders and items and derive optimal prices\footnote{Vickrey-Dutch Auctions are multi-item, sealed-bid auctions that maximise a measure of \emph{social welfare} for their assignments (\ie sum of the differences between initial bids and paid prices). VDA assigns the price of the \emph{second highest} bid to each item, incentivising truthful valuations.}, and can be easily adapted to support other forms of auctions.
\sysname extends VDA with support for general bids: Bidders can provide a valuation for any item whose description matches a set of predicates.
General bids are automatically transformed in a number of per-item bids, as manipulated by the VDA algorithm.

\sysname leverages the Ethereum blockchain for transparency, enabling any party to audit and if necessary invalidate the outcome of an auction.
To ensure that bidders only provide faithful valuations, \sysname binds bids to the Ethereum cryptocurrency: Buyers cannot bid higher than the amount of money submitted to a deposit.
Funds are automatically deducted from their accounts if they acquire an item. 

\sysname protects privacy by hiding submitted bids as well as the identity of bidders using threshold blind signatures.
In contrast with related work, \sysname does not rely on a single trusted third party, but rather on a set of multiple, decentralised authorities.
Each user can define their own set of trusted parties and is protected from a definable subset of authorities becoming malicious.

High efficiency in \sysname results from two key design choices.
First, we run the heavy VDA computation off-chain and allow any node to submit an auction solution in the form of an assignment between items and bidders and a price vector.
Second, to avoid re-running auctions or perform costly on-chain verification we develop simple verification techniques in the form of \emph{proofs of misbehaviour}.
Users can generate such proofs without re-running the VDA algorithm, and for a fraction of its computational cost.
The \sysname smart contract can check the validity of these proofs and discard faulty assignments.
Both providers of solutions and submitters of proofs of misbehaviour are held accountable for their actions by linking virtuous and malicious behaviours with monetary rewards and penalties.

We implement \sysname and deploy it over Ethereum, using workloads derived from Filecoin.
\sysname ensures correctness, protects bidder's privacy and scale to large of numbers bidders and items. Our evaluation shows the ability to derive optimal assignments between 10,000 users and 100 Filecoin storage nodes with less than 8 seconds of CPU time on a regular laptop and enable auditing the result in less than half a second. The comparison with Verifiable Sealed-bid Auctions~\cite{galal2018verifiable} shows 4 orders of magnitude lower execution time, 2.5x reduction in contract deployment costs, and up to an 8x reduction in cost per bidder. \sysname source code is publicly available for the scientific community and easy to adapt to support any sharing-economy platform\cite{pastrami}. 

\para{Outline}
The rest of this paper is structured as follows.
\Cref{sec:goals} presents our system, its adversary model, and details our objectives.
\Cref{sec:overview} presents an overview of \sysname, and \Cref{sec:background} discusses background.
We describe our design divided into a preparation phase and an execution phase in Sections~\ref{sec:preparation} and~\ref{sec:execution}.
We provide a discussion and a security analysis in \Cref{sec:analysis}, and present our evaluation results in \Cref{sec:evaluation}.
We finally cover related work in \Cref{sec:related} and conclude in \Cref{sec:conclusion}.

%% file: sections/goals.tex


\section{\sysname Design Goals}
\label{sec:goals}

We start by defining our system model, notations and assumptions. 
We follow up by specifying our target properties for correctness, but also privacy and ease of deployment.
These properties are summarised in \Cref{tab:properties}.
We describe system components using Filecoin as a running example.
However, our platform can be used in any decentralised system requiring an assignment between buyers and sellers~\cite{golem,iexec,lavoie2018pando,sonm,storj}.

\subsection{System Model}

We consider the following actors: 
\begin{itemize}
\item \textbf{Bidders} are users wishing to buy goods or services, denoted by the generic term \emph{item}, e.g., Filecoin end-users willing to store their files;
\item \textbf{Sellers} are users offering such items, e.g., Filecoin storage nodes;
\item A set of \textbf{authorities} jointly issues credentials allowing bidders to anonymously participate in auctions. For instance, in Filecoin this role can be taken by blockchain validators or a distinctive set of entities trusted by system participants.
\end{itemize}

We emphasise that none of the authorities is trusted individually by the users.
Instead, users trust a set of authorities \emph{as a whole} similarly to an honest majority in blockchains or Byzantine fault-tolerant protocols~\cite{castro1999practical}.

\sysname accepts item description submitted by sellers and sealed-bids commitments (\emph{bids}) expressed by bidders.
A bid expresses a \emph{valuation}, \ie the maximal price that the bidder is willing to pay for an item.
Each bid is backed by a deposit of that valuation.
This deposit is expressed in coins that cannot be double-spent~\cite{karame2012double} and is guaranteed by the authorities. 
Based on the set of all bids, \sysname outputs a price vector and an assignment between bidders and items.
This assignment must be the output of a globally-known algorithm.
\sysname employs the Vickrey-Dutch multi-item auction algorithm, but can be adapted to different types of single-item, multi-item and even combinatorial auctions.

In the example of Filecoin, storage nodes (sellers) advertise their capacities (\ie storage space, available bandwidth, lease duration) together with a minimum price they are willing to accept.
End-users (buyers) submit their requirements and a maximum amount they are willing to pay (\ie ``I am willing to pay 10 tokens for a month of using a 80~GB, high-bandwidth storage node'').
\sysname automatically allocates end-users to the most suitable storage nodes and derives prices following rules of supply and demand.

\subsection{Notations and Assumptions}
\label{sec:notations}



\para{Cryptographic assumptions}
\sysname inherits the same cryptographic assumptions as its underlying BLS blind signature scheme we present in \Cref{sec:background}, which requires groups $(\mathbb{G}_1,\mathbb{G}_2,\mathbb{G}_T)$ of prime order $p$ with a bilinear map $e:\mathbb{G}_1 \times \mathbb{G}_2 \rightarrow \mathbb{G}_T$ and satisfying \first\emph{Bilinearity}, \second\emph{Non-degeneracy}, and \third\emph{Efficiency}.
\sysname uses type-3 pairings because of their efficiency~\cite{galbraith2008pairings}; and thus relies on the XDH assumption which implies the difficulty of the Computational co-Diffie-Hellman (co-CDH) problem in $\mathbb{G}_1$ and $\mathbb{G}_2$, and the difficulty of the Decisional Diffie-Hellman (DDH) problem in $\mathbb{G}_1$~\cite{bls}.

\para{Threshold assumptions}
\sysname assumes that at least $t$-out-of-$n$ authorities are honest and available at all time (with $t > n/2$).
If more than $t$ authorities collude, they can forge signatures or stall the system, but cannot break unlinkability and de-anonymize users.

\para{Communication assumptions}
The \sysname actors interact through the direct exchange of messages as well as through a blockchain supporting executable \emph{smart contracts}.
Our implementation uses Ethereum~\cite{buterin2013ethereum}.
We assume the following about communications:
\begin{itemize}[leftmargin=9pt]
    \item \textbf{Authorities:} \sysname authorities are collectively responsible for issuing signatures to bidders. 
	They do not need to communicate with each other in order to issue these signatures. 
	However, practical implementations of threshold signature schemes initially require either using the distributed key generation protocol of Kate~\etal~\cite{kate2012distributed}, which requires \first weak synchrony for liveness (but not for safety), and \second at most one third of dishonest authorities; or using the distributed key generation protocol of Gennaro~\etal~\cite{gennaro1999secure} which requires \first synchrony, and \second an honest majority.
	Key generation is, however, run only once (or rarely) and resulting keys can be used in multiple auctions.
    \item \textbf{Authorities-Users:} Users wait for $t$-out-of-$n$ replies (in any order of arrival) and aggregate them into a consolidated signature; thus \sysname implicitly assumes an asynchronous setting between users and authorities.
    \item \textbf{Bidders-Sellers:} Bidders do not interact or trust each other. The same applies to sellers. A bidders and a seller need only communicate directly to implement a deal, after an auction final assignment is available.
\end{itemize}

\begin{table}[t!]
\resizebox{\linewidth}{!}{%
\footnotesize

\newcolumntype{C}{>{\raggedright\let\newline\\\arraybackslash\hspace{0pt}}m{0.31\linewidth} }
\newcolumntype{D}{>{\raggedright\arraybackslash} m{0.725\linewidth} }

\begin{tabular}{CD}
\toprule

\multicolumn{2}{l}{\textbf{Economic Properties}}                          \\ \midrule
Efficiency                    & bidders get items they value the most \\
Incentive Compatibility       & bidders benefit from revealing their bids\\
Individual Rationality        & bidders and sellers benefit from participation\\
Budget Balance                & derived prices exceed minimum prices\\
\midrule
\multicolumn{2}{l}{\textbf{Correctness}}                        \\ \midrule
Bids Binding                  & bids cannot be changed once committed\\
Public Auditability           & participants are able to verify its correctness\\
Fairness                      & users deviating from the protocol are penalised\\ \midrule
\multicolumn{2}{l}{\textbf{Privacy}}                \\ \midrule
Hidden Minimum Price          & minimum prices are private until the end of the auction\\
Bidders' Privacy               & bidders are unlinkable to their bids \\
Bids Privacy                  & bids are private until the end of the auction\\ \midrule
\multicolumn{2}{l}{\textbf{Deployment}}                          \\ \midrule
Non-Interactivity             & bidders are not required to stay online \\
Openness                      & anyone can become a bidder or a seller \\
Distributed authority         & \sysname does not rely on a single trusted party \\
Scalability                   & support for large number of items and bidders \\
\bottomrule
\end{tabular}
}
\caption{Target properties of \sysname.}
\label{tab:properties}
\end{table}

\subsection{Target Properties for \sysname}
\sysname aims at providing the following properties:

\para{Auctions' economics} 
\sysname assigns resources to the \bidders that value them the most guaranteeing \emph{Efficiency}. 
In addition, \sysname guarantees \emph{Incentive Compatibility} meaning that \bidders and \sellers benefit from revealing their true valuations and can only lose by artificially increasing or decreasing those valuations.
This property is linked to \emph{Individual Rationality}, which ensures that both \bidders and \sellers are incentivised to participate in the auctions. 
Finally, it targets \emph{Budget Balance} by making sure that the payments submitted cover \sellers' compensations (\ie derived prices are superior to minimum prices submitted by sellers).

\para{Correctness} 
\sysname verifies that auction assignment are computed correctly and that all participants follow the protocol.
To avoid malicious manipulation, \bidders cannot change their bids once they are committed (\emph{Bids Binding}).
Furthermore, anyone can verify the correct execution of any auction by inspecting the public information available on the ledger (\emph{Public Auditability}).
\sysname achieves \emph{Fairness} by making sure that \bidders are financially penalised if they deviate from the protocol, but cannot be financially penalised if they follow it correctly.

\para{Privacy}
\sysname aims to only disclose the minimum amount of information required to verify the correctness of the auction.
The minimum prices submitted by sellers are kept private from the \bidders until the end of the auction (\emph{Hidden Minimum Price}); and bids submitted to the system are kept private until the end of the auction (\emph{Bids Privacy}).
Those two properties ensure that \bidders submit their bids without any knowledge of what others do, preventing price manipulation (\ie submitting bids slightly above the minimum price or the current highest bid).
\sysname also keeps \bidders anonymous, ensuring that \bidders are unlinkable to their bids and that the identity of the assigned bidder is revealed only to the \seller at the end of the auction.

\para{Deployment}
The process of submitting items, bids and calculating a result can take a significant amount of time.
Many participants may not be able to stay connected during the entire duration of an auction.
To facilitate large scale deployments and support large number of users, \bidders are not required to interact with each other providing \emph{Non-Interactivity}.
Furthermore, \sysname targets \emph{Openness}, meaning that anyone can act as \bidder or \seller and the system is resilient to censorship.
Our platform support the \emph{Distributed Authority} property and does not rely on a single trusted 3$^{\text{rd}}$ party at any point.
Finally, \sysname achieves high \emph{Scalability} and supports large number of users and items while maintaining low overall cost of running auctions for the platform and its participant.

%% file: sections/overview.tex
\section{\sysname Overview}
\label{sec:overview}

\begin{figure}[t]
\includegraphics[width=\linewidth]{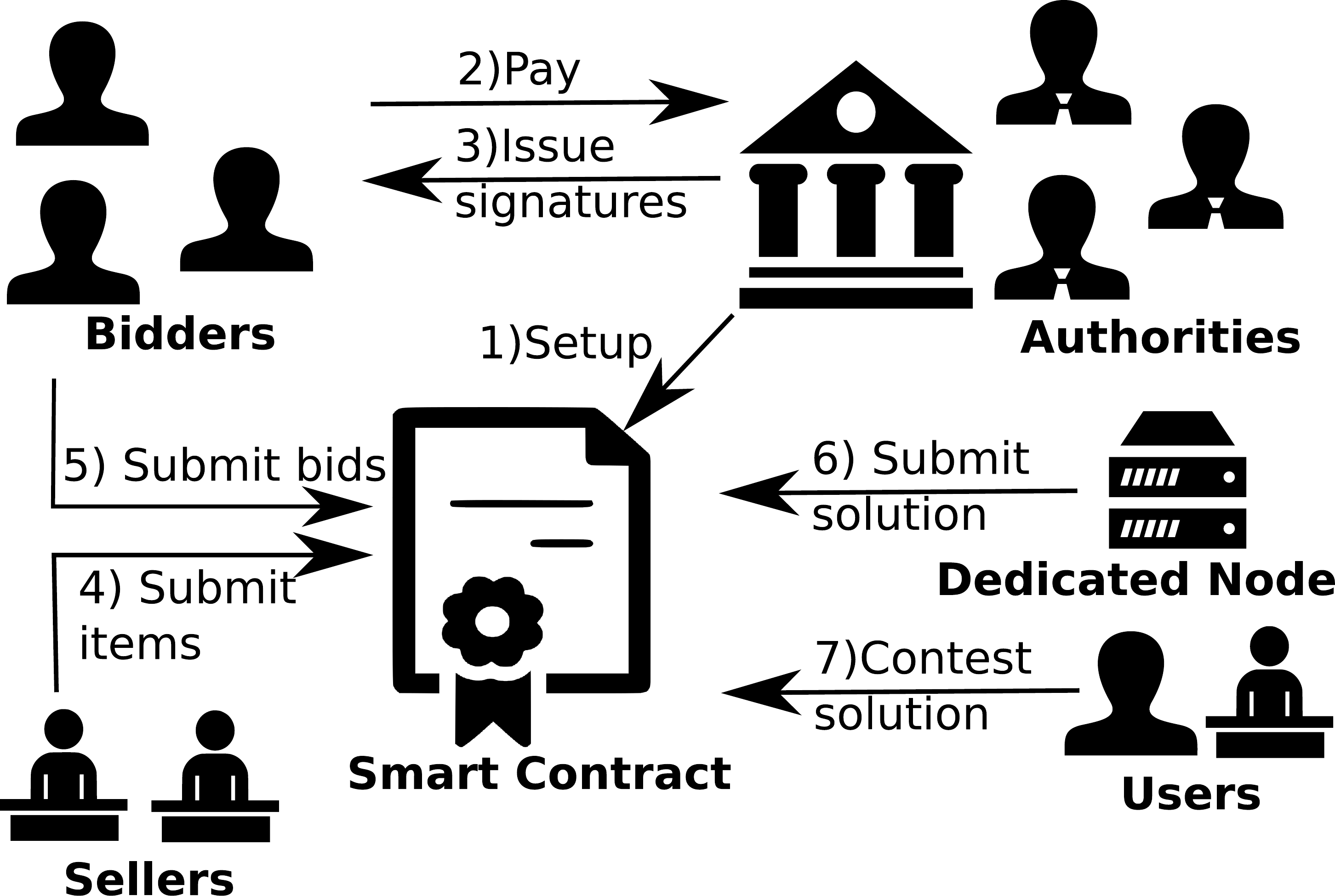}
\caption{\sysname overview.}
\label{fig:overview_new}
\end{figure}

\Cref{fig:overview_new} presents an overview of \sysname.
An auction is divided in two subsequent phases, a \emph{preparation phase} (steps~1 to~5) and an \emph{execution phase} (steps 6 and 7).

\para{Preparation phase}
A set of distributed authorities starts by publishing their aggregated verification key along with any associated parameters, and sets up a smart contract on the blockchain (step~1).
In order to create privacy-preserving accounts, \bidders contact this smart contract and pay a maximum amount they are willing to bid as a deposit (step~2). 
The authorities observe the smart contract, and generate cryptographic material required by \bidders to participate in the auction.
\Bidders then contact each authority, retrieve parts of the material and combine these parts locally into cryptographic credentials (step~3).
Those credentials allow \bidders to participate in auctions while guaranteeing strong privacy and ensuring that bidders are backed by a deposit.
Bids remain unlinkable to \bidders's real identities or to cryptographic material issued by the authorities. 

An auction begins with \sellers submitting descriptions of their items/services to \sysname. 
To guarantee that their activities are profitable, \sellers can specify a hidden minimum price for which they are willing to trade their items, and are guaranteed that the users assigned during the auction have bid for at least that price (step~4).

In step~5, bidders submit their sealed bids to the smart contracts using their credentials. 
Bidders can submit two types of bids.
Regular bids express the interest of the bidder for a specific item, in a form of a price \emph{valuation}.
\emph{General} bids allow, on the other hand, bidders to specify their interest in \emph{any} one of the items satisfying some constraints.
Auction algorithms only support single-item evaluation, henceforth \sysname automatically derives a vector of single-item valuations for all items matching the constraints set in a general bid.
Once all the information is submitted to the blockchain, bids\footnote{The revealed bids are unlinkable to bidders who expressed them.} and minimum prices are revealed.

\para{Execution phase}
Instead of performing costly assignment calculations on-chain, the smart contract allows a dedicated node\footnote{We use a dedicated node for simplicity.
In practice, any node can calculate a solution and submit it to the chain.} to use the information stored on the blockchain to perform computations off-chain and submit a solution to the platform (step 6).
The calculation of a solution off-chain does not suffer from the overhead associated with distributed and privacy-preserving solutions such as MPC (\Cref{sec:related}).
A solution contains an assignment between sellers and bidders and derived prices for each of the assigned items.
A score is associated to this assignment, representing the \emph{social welfare}, a measure of the quality of the compromise between buyers and sellers' interests.

Once submitted, the solution can be contested within a specified amount of time.
Any user, bidder or seller, may verify the solution off-chain and submit a proof of misbehaviour (step 7).
\sysname uses lightweight verification that is much faster than recomputing the auction result.
The smart contract can efficiently test the validity of each proof of misbehaviour and reject the contested solution.
Alternatively, if no valid proof of misbehaviour is received by the end of the time period, the solution is marked as final.
Bidders can contact their corresponding sellers, prove their identity and claim acquired items. 

In the next sections, we start by detailing the background constructions used by \sysname, followed by the details of the preparation and execution phases.

%% file: sections/background.tex


\section{Background}
\label{sec:background}

\para{Smart Contracts on Blockchains}
The concept of a blockchain was introduced by Bitcoin~\cite{bitcoin} as a decentralised, append-only ledger that provides a global ordering of financial transactions, to prevent funds being spent twice (the \emph{double-spending} attack). 

Blockchain platforms such as Ethereum~\cite{buterin2013ethereum} provide script languages to allow users to execute more complex programs on the blockchain, called \emph{smart contracts}.
Executing a transaction calling a method of an Ethereum smart contract has an associated \emph{gas} cost that is proportional to the number of instructions and the amount of storage required by the transaction.
The monetary value of gas varies depending on the load on the network, and is paid for using Ether, Ethereum's built-in currency.

\sysname leverages the high-integrity data structure provided by a blockchain, and uses it for accountancy, auditability, and time reference (\eg time is defined as a difference in block height in the chain).

\para{Threshold Blind Signatures}
Blind signatures are digital signatures allowing users to hide (blind) the content of a message from the signer.
The user can then locally unblind the signature, which can then be publicly verified as traditional digital signatures~\cite{chaum1983blind}.
Most blind signature schemes entrust a \emph{single} authority with a master signing key, allowing a malicious authority to forge any signature.
To overcome this limitation, \sysname relies on \emph{threshold blind signatures}~\cite{bgls} that distribute the signing process between multiple authorities.  

We give below the high-level definition of a threshold blind signature scheme.
For the sake of simplicity, we use in this description a key generation algorithm \algorithm{BS.TTPKeyGen} as executed by a single trusted third party.
This protocol can, however, be executed in a distributed way as illustrated by Gennaro~\etal~\cite{gennaro1999secure} under a synchrony assumption, and as illustrated by Kate~\etal~\cite{cryptoeprint:2012:377} under a weak synchrony assumption.
All algorithms receive the security parameter $\lambda$ as an input but we show it explicitly only for \textsf{Setup}. \Cref{sec:blind-bls} provides the full cryptographic construction of the blind BLS signature scheme.

\begin{description}[leftmargin=1em, labelindent=0em]
\setlength\itemsep{0.5em}
\item[\definition{BS.Setup($1^\lambda$)}{$\mathit{params}$}] defines the system parameters $\mathit{params}$ with respect to the security parameter $\lambda$ (these parameters are publicly available).

\item[\definition{BS.TTPKeyGen($params, t, n$)}{$x, y$}] run by the authorities to generate a secret key $x$ and a verification key $y$ from the public parameters $params$.

\item[\definition{BS.PrepareBlindSign($m$)}{$\tilde{h}$}] run by the user to request a blind signature over the cryptographic material $\tilde{h}$ embedding the message $m$.

\item[\definition{BS.BlindSign($x_i, \tilde{h}$)}{$\tilde{\sigma}_i$}] run by each authority $i$ to issue a partial blind signature $\tilde{\sigma}_i$ (using their private key $x_i$) over the user-provided cryptographic material $\tilde{h}$.

\item[\definition{BS.Unblind($r, \tilde{\sigma}_i$)}{$\sigma_i$}] run by the user to unblind the signature $\tilde{\sigma}_i$ and recover the signature $\sigma_i$ (using the blinding factor $r$). 

\item[\definition{BS.AggSig($\sigma_1, \dots, \sigma_t$)}{$\sigma$}] run by the user to aggregate $t$ partial signatures $(\sigma_1, \dots, \sigma_t)$ into a consolidated signature $\sigma$.

\item[\definition{BS.Verify($y, m, \sigma$)}{$\mathit{true}/\mathit{false}$}] run by any third party verifier to check the validity of a signature $\sigma$ over the message $m$ (using the aggregated public key $y$).
\end{description}


\subsection{Multi-item Auction Execution}
\label{sec:auction}

We consider a single smart contract that sets up an auction for a set of $\mathcal{I}=\{1,2,...,I\}$ items.
A set of $\mathcal{B}=\{1,2,...,B\}$ bidders are interested into the offered items of the contract where each bidder $b\in\mathcal{B}$ is expressing their preference for each item $i\in\mathcal{I}$ via a \textit{private monetary valuation}, denoted by $v_{bi}\geq 0$.
We focus on a \textit{unit-demand} setting where each bidder is interested in acquiring \textit{at most a single} item\footnote{A bidder interested in buying multiple items may participate multiple times to the same auction under different, un-linkable identities.}.
That is, we consider a \textit{null item}, denoted by $0$, that is assigned to the bidders who fail to acquire an item in the auction.
Note that the valuation of each bidder for the null item is zero, \ie $v_{b0}=0$ $\forall b\in\mathcal{B}$.
For notational convenience let $\mathcal{I}^*=\mathcal{I}\cup\{0\}$.

A \textit{feasible} allocation $X$ assigns an item $x_b\in\mathcal{I}^*$ to each bidder $b\in\mathcal{B}$ such that for $b\neq b'$ $x_b\cap x_{b'}=\emptyset$ or $x_b\cap x_{b'}=0$, meaning that only the null item 0 can be allocated to more than one bidder.
A feasible allocation is consider \textit{efficient}, denoted by $X^*$, if there is no allocation $X$ such that $\sum_{b\in\mathcal{B}}v_{b{x_b}}>\sum_{b\in\mathcal{B}}v_{b{x^*_b}}$.

Auctions are pricing mechanisms that lead to a feasible allocation of items.
In detail, an auction associates each item $i\in\mathcal{I}$ to a price $p_i$ forming the \textit{price vector} $p=(p_0,p_1,....,p_I)$.
The seller of an item $i$ can set a \textit{reservation price} $r_i\geq 0$ that is defined as the minimum price that it is willing to offer the item for, restricting the auction's assigned prices for the specific item to $p_i\geq r_i$.
Given a price vector $p$, the interest of each bidder is limited to the items that return the highest valuation after reducing the corresponding price.
That is, the demand correspondence $D_b(p)$ for bidder $b\in\mathcal{B}$ is given by $D_b(p) = \{i\in\mathcal{I}^*:v_{bi}-p_i\geq v_{bj}-p_j \forall j\in\mathcal{I}^*\}$.

A price vector $p^*$ is an \textit{equilibrium price vector} if it derives a feasible assignment $X$ where an item $i$ remains unassigned at price $p_i^*=r_i$ or it is uniquely assigned to a bidder $b\in\mathcal{B}$, \ie $x_b=i$ and $x_b\neq x_{b'} \forall b'\in\mathcal{B}\backslash\{b\}$, where $x_b\in D_b(p)$ at price $p_i^*\geq r_i$.
That is, the pair $(X,p^*)$ is an \textit{equilibrium allocation} if $p^*$ is an equilibrium price vector.
The set of competitive price vectors is non-empty and forms a complete lattice~\cite{shapley1971assignment}, meaning that there is a unique minimal element that is referred as the Vickrey-Clarke-Groves (VCG) price vector and is denoted by $p^\textrm{VCG}$.
Furthermore, the corresponding assignment to $p^\textrm{VCG}$ price vector is efficient, denoted by $X^\textrm{VCG}$.

This result essentially states that when the price vector $p^\textrm{VCG}$ is applied, an item is assigned to a user at the lowest possible price over any possible equilibrium price vector, \ie any feasible assignment that respects bidders' demand correspondence.
In other words, there is no other equilibrium than VCG that bidders would prefer, since it satisfies their demand in a stable way at the lowest possible price by deriving the highest possible \textit{net} valuations (\ie the difference between the bid and the obtained price).
This has the implication that under an auction that assigns items according to VCG equilibrium, bidders have no incentives to interfere with the smooth operation of the mechanism; therefore is to the best of their interest to be \textit{truthful} upon the declaration of valuations for each item of the auction.

There are plenty of auction mechanisms that derive the VCG equilibrium in polynomial time.
We use the Multi-Item Unit Demand Vickrey-Dutch Auction (VDA)~\cite{mishra2004multi,mishra2009multi} as the underlying auction of \sysname\footnote{Alternative mechanisms include the Vickrey-English and Vickrey-English-Dutch auctions which are also polynomial although are coming at a higher complexity compared to VDA.}.

%% file: sections/preparation.tex
\section{Preparation Phase}
\label{sec:preparation}

We present the protocol behind the preparation phase followed by a formal description of the algorithms.

\subsection{Preparation Phase Protocol}
\para{Setup}\label{sec:setup}
A set of authorities execute the \algorithm{Setup} method of the \sysname smart contract; they provide their public keys as well as any other scheme parameters (or policy) as the number of \authorities and the threshold parameter; publish a unique identifier for the auction $\mathit{id}$, and initiate a timer $T$ (expressed in a number of blocks). 
Once created, any \seller can add a description of an item to be sold and a commitment to a minimum price $r_i$.

\para{Commit}\label{sec:commit}
\Bidders execute \algorithm{Commit} by paying a deposit of $d$ coins to the \sysname smart contract, and specifying cryptographic material embedding a fresh account address $\mathit{addr}$ they own, the auction unique identifier $\mathit{id}$, a random number $k$, and their bid $b$.
The contract emits an event instructing the authorities to issue a blind signature using this cryptographic material.
The value $d$ represents the number of coins the bidder wishes to bid.
To mitigate traffic analysis, $d$ should be limited to a specific set of possible values, similar to cash denominations.
Each authority monitors the \sysname smart contract, and issues a partial blind signature to the user---either on chain or off-chain---upon detecting the request (signature requests are processed only if \bidders paid a deposit of $d$ coins to the smart contract and correctly executed \algorithm{Commit}).
Authorities use a different set of keys for each cash denomination (\eg if the user deposits $d=5$, the authorities issue the blind signature using a key pair $(sk_5, pk_5)$).
\Bidders locally unblind and aggregate all partial signatures into a consolidated signature.
All account addresses in this protocol must be fresh in order to not leak information about the identity of the users; users can privately add coins to an address using a coin tumbler~\cite{mixcoin,blindcoin,tumblebit,coinjoin,coinshuffle,xim,mobius, coconut}.

\para{Reveal}\label{sec:reveal}
After the timer expires, \sellers open the commitment to their minimum price. 
\Bidders execute \algorithm{Reveal} and submit their unblinded signatures to the contract using their fresh account address $\mathit{addr}$.
The contract accepts the bid only if it is submitted by the same address $\mathit{addr}$ as the one embedded in the signature, if the auction $\mathit{id}$ matches the current auction, and if it has not already seen the random number $k$; the contract keeps track of all the numbers $k$ it has seen thus far to prevent double-spending.
The contract interprets the value associated with each signature based on their signing key.
Including $addr$ in the blind signature and verifying that it is used to submit the bid prevents malicious actors from intercepting the bid during \textsf{Reveal} and spending it on behalf of victim users.

\subsection{Algorithm Constructions}
\label{sec:algorithms}

We present our algorithms.
\textsf{BS}-prefixed algorithms are from the threshold blind BLS signature scheme (\Cref{sec:background}). 

\para{Interactions between users and smart contract}
We first describe the algorithms allowing users to interact with the \sysname smart contract.
We prefix bidders' functions (executed off-chain by the client's software) with a capital $B$, and smart contract functions (executed on-chain by the \sysname smart contract) with a capital $C$.
All algorithms have implicitly access to the address identifying the account of the bidder who sends the transaction. 
Symbol $||$ denotes string concatenation.

\begin{description}[leftmargin=1em, labelindent=0em]
\setlength\itemsep{0.5em}
\item[\definition{Setup($1^\lambda$)}{$\mathit{params}$}] \ \ 
\\ $\triangleright$ Defines the system parameters $params$ with respect to the security parameter $\lambda$ (these parameters are public). \\
Output \textsf{BS.Setup($1^\lambda$)}; the smart contract creates an empty spent-list $\mathcal{L}$ in memory, and initializes a timer $T$. 

\item[\definition{Commit($d, \mathit{addr}, \mathit{id}, k, b, T$)}{$ $}] \ \
\\ $\triangleright$ Can only be executed if the timer $T$ has not expired; commit to the bid $b$ along with the parameters $\mathit{addr}, \mathit{id}, k$ by sending $d$ coins to the contract.
\begin{description}
\item\definition{B.Commit($d, \mathit{addr}, \mathit{id}, k, b$)}{$\tilde{h}$} \\ Compute $m = (\mathit{addr}||\mathit{id}||k||b)$, \\ and $\tilde{h}=\textsf{BS.PrepareBlindSign($m$)}$; send $\tilde{h}$ along with $d$ coins to the appropriate smart contract instance.
\item\definition{C.Commit($\tilde{h}, T$)}{$ $} if the timer $T$ has not expired, emit an event telling the authorities to issue a blind signature to the bidder; otherwise ignore.
\end{description}

\item[\definition{Reveal($m, \sigma, y, T$)}{$ $}] \ \
\\ $\triangleright$ Reveal the bid by revealing the unblinded signature.
\begin{description}
\item\definition{B.Reveal($m, \sigma$)}{$ $} submit $(m, \sigma)$ to the smart contract using the address $\mathit{addr}$.
\item\definition{C.Reveal($m, \sigma, y$)}{$ $} parse $m = (\mathit{addr}||\mathit{id}||k||b)$;  save the bid $b$ in memory if and only if the sender address is $\mathit{addr}$, the auction $\mathit{id}$ matches the current auction, $k \notin \mathcal{L}$, and if $\textsf{BS.Verify($y, m, \sigma$)}=\mathit{true}$; then append $k$ to $\mathcal{L}$. Otherwise ignore.
\end{description}
\end{description}

\para{Interactions between bidders and authorities} 
We describe the algorithms allowing users to interact with the authorities.
We prefix off-chain bidders' functions with a capital $B$, and off-chain authority functions with a capital $A$.
 
\begin{description}[leftmargin=1em, labelindent=0em]
\item[\definition{Issue($sk, \tilde{h}$)}{$\tilde{\sigma}$}]\ \
\\ $\triangleright$ The authorities issue a blind signature to the user over the cryptographic material $\tilde{\sigma}$.
\begin{description}
\item\definition{A.Issue($sk, \tilde{h}$)}{$\tilde{\sigma}_i$} each authority returns to the user $(\tilde{\sigma}_i)$ = \textsf{BS.BlindSign($sk, \tilde{h}$)}.
\item\definition{B.Issue($\tilde{\sigma}_1, \dots, \tilde{\sigma}_t$)}{$\sigma$} for each $\tilde{\sigma_i}$, compute $\tilde{\sigma}$ = \textsf{BS.AggCred($\tilde{\sigma_1}, \dots, \tilde{\sigma_t}$)}; output $(\sigma)$ = \textsf{BS.Unblind($r, \tilde{\sigma}$)}.
\end{description}
\end{description}


\subsection{Expressing valuations}
\label{sec:valuations}
Regular valuations are submitted as price vectors without any additional support. 
To enable general bids, we enhance item descriptions submitted by sellers by a set of characteristics $\mathcal{C} = \{1,2,...,C\}$.
Each characteristic represents a single property of the submitted item.
In the example of FileCoin, these properties could be the storage size, the service duration, or the seller reputation.
For simplicity, let all characteristics be expressed as integers. 

A general bid uses a set of constraints for each characteristic $f_{b}$ and a maximum budget $d_{b}$ instead of a private monetary valuation for each item $v_{bi}$.
Each bidder is interested uniquely in items fulfilling all the submitted constraints. Furthermore, we use the constraint vector to automatically derive private monetary valuations required by the auction algorithm $\textit{derive}(f_{b}) = v_{bi}$ (see \Cref{sec:auction}).

The constraint vector of a general bid, while less flexible than single item valuation allows to radically decrease the amount of information submitted to the blockchain and enhance bidders' privacy.
Finally, the constraint vector is reusable and can be submitted without knowing the list of offered items, reducing further the overhead of our solution.

%% file: sections/execution.tex
\section{Execution Phase}
\label{sec:execution}

\begin{figure}[b]
\includegraphics[width=\linewidth]{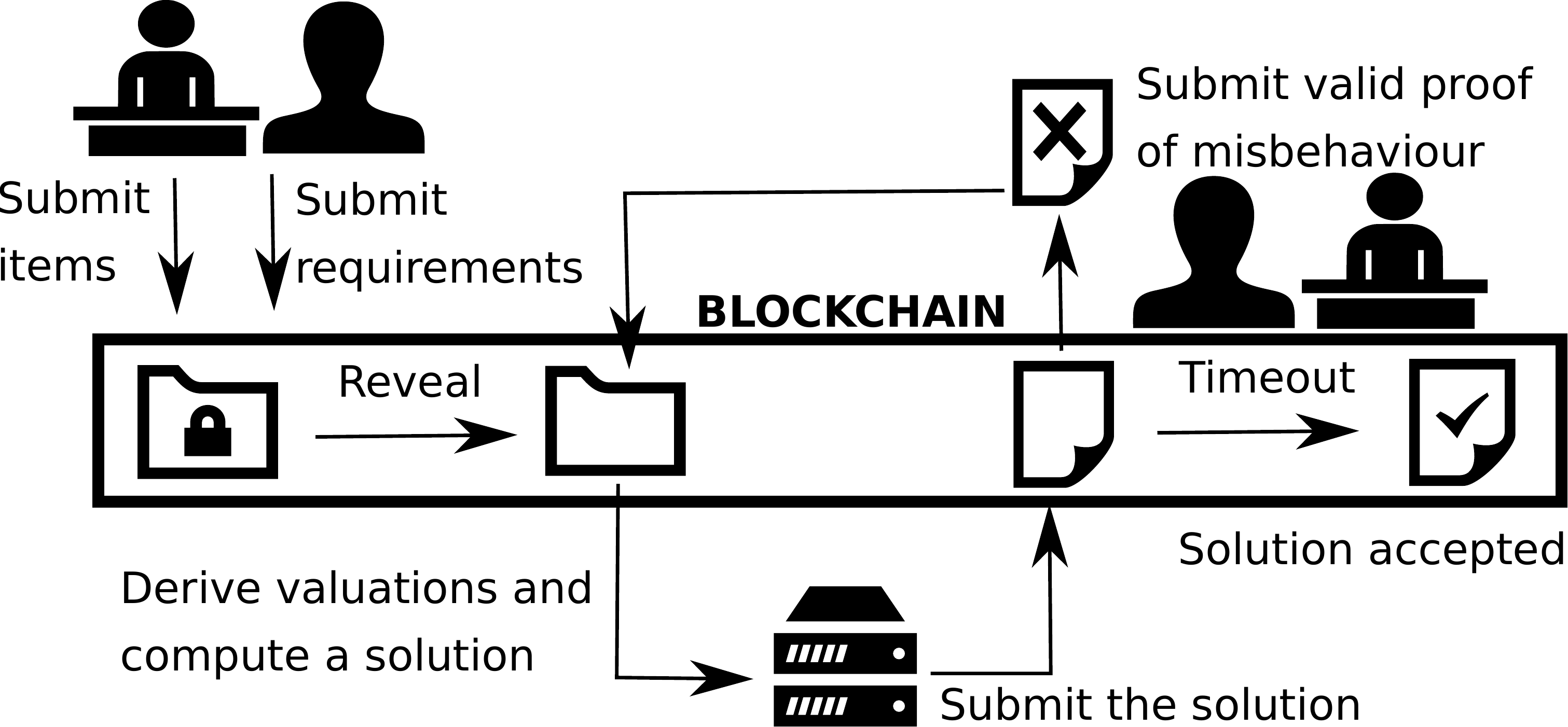}
\caption{The smart contract offloads heavy computations and accepts computed solutions that can be contested by other users.}
\label{fig:offloading}
\end{figure}


At this time, all the information to determine an assignment is available on the blockchain.
This assignment is calculated by the dedicated node.
The first step is to derive bidders' valuations from the \emph{general} bids revealed in \algorithm{Reveal} (\Cref{fig:offloading}).
The derivation function assigns the specified maximum budget $v$ to all the items fulfilling the requirements $v_{bi} = v$.
If at least one characteristic is lower than a specified constraint the assigned valuation is set to 0, \ie $v_{bi} = 0$.
Once valuation vectors are computed, the dedicated node uses an unmodified version of the Vickrey-Dutch auction algorithm~\cite{mishra2004multi,mishra2009multi}.

In order to incentivise correct behaviour, the dedicated node must first provide a collateral $d_{s}$.
This collateral is returned once the solution is marked as final.
We calculate the minimum amount of the deposit in \Cref{sec:evaluation}.

A submitted solution consists of an assignment of items to bidders $X$, a price vector $p$ and a score $s$.
$p$ indicates the established price of each item, while $s$ represents the social welfare score (\ie the sum of net valuations).
By definition, VCG equilibrium dictates the solution with the highest social welfare, where all items are universally allocated (\Cref{sec:background}).

Once a solution is stored on-chain, bidders and sellers have a predetermined amount of time $t$ to verify it and submit a proof of misbehaviour.
If none is received after time $t$ the solution is marked as final.



\subsection{Verification and Proofs of Misbehaviour}
\label{sec:proofs}

Given an assignment $X$, a price vector $p$ and a score $s$, an auction solution is correct if:
\begin{itemize}[leftmargin=9pt]
 \item every user is satisfied with their assigned item at a given price, \ie the solution is an equilibrium;
 \item no price is lower than the reservation price;
 \item the declared score is correct;
 \item every bidder pays for their assigned item the minimum possible price over all possible equilibriums, which characterises the VCG equilibrium.
\end{itemize}

\para{Equilibrium misbehaviour}
A user verifies whether the bidders got assigned the most optimal items (with the highest net valuation, \ie bidder's valuation after subtracting the item's price).
Let the assigned item for a bidder $b\in\mathcal{B}$ be $x_{b} = i$.
For each bidder we have to check if there is another item $\exists j\in\mathcal{I}\backslash\{i\}$ that the bidder would prefer at its current price, \ie $v_{bi} - p_{i} < v_{bj} - p_{j}$; if it is the case, it is an indication that the proposed assignment is not an equilibrium.
In a nutshell, the verification consists of a single optimization step of the VDA algorithm. 
The process detects all the possible misconducts and uses a fraction of the cost required to compute the correct solution. 
Users who detect the existence of such an item can submit a proof of misbehaviour using the \algorithm{wrongAssignment} smart contract method, pointing out the bidder $b$ and the alternative item $j$ associated to a higher net valuation.
Upon reception of this proof the smart contract derives bidder's $b$ current net valuation, $v_{bi} - p_{i}$ as well as the proposed alternative net valuation $v_{bj} - p_{j}$ before comparing the two.
The entire operation requires only 2 subtractions and a single comparison and therefore its monetary cost is negligible, as we show in \Cref{sec:evaluation}. 

\para{Price misbehaviour}
A user proceeds by verifying whether items are offered at a valid price, \ie greater than their reservation price.
If it is not the case, any user can point out an index in $p$ with a price lower than the reservation price using the \algorithm{wrongPrice} method. 
The whole operation requires just a single comparison executed on the smart contract.

\para{Score misbehaviour}
Each candidate solution contains a score $s$, corresponding to the sum of net valuations.
Users checks its validity in terms of $\sum_{b\in\mathcal{B}}v_{bx_{b}} - p_{x_{b}}= s$.
This is done by iterating through the assigned items and their price vector while computing the sum of net valuations.
If the score is incorrect a user invokes \algorithm{wrongScore}. 

To verify the proof, the smart contract has to compute the score on-chain, which may be expensive.
However, the cost for this calculation will reimbursed by the collateral submitted by the dedicated node if the score was indeed incorrect.

\para{VCG equilibrium misbehaviour}
So far, we explained how to verify that a candidate solution is an equilibrium, but not that it is the VCG equilibrium.
A VCG equilibrium is defined as the one maximising the sum of bidders' net valuations, \ie $\sum_{b\in\mathcal{B}}v_{bx_{b}} - p_{x_{b}}$.
Users compute a set of excess demand and supply.
This procedure consists of running a \emph{single} iteration of the Vickrey-Dutch algorithm and can be efficiently performed in polynomial time (\Cref{fig:filecoin_cpu}).
A non-empty set indicates that the solution is not a VCG equilibrium with the highest possible score and is thus incorrect.

\para{Proofs submissions timescale} 
If no valid proof of misbehaviour is submitted by the deadline, the solution is marked as final and the auction finishes.
The majority of blockchain platforms do not support scheduling invocation in the future. However, it can be realised by a public function that checks if enough time has passed before marking the solution as final.
The submitted solution indicates items assigned to each bidder together with the price to pay that is lower than $v$.
\Bidders that correctly followed the protocol but get no assigned items may withdraw all of their deposited coins by calling the \algorithm{Withdraw} function of the \sysname contract.
Recall that \sysname applies a Vickrey-Dutch auction mechanism~\cite{vickrey1961counterspeculation}; the winner of the auction is the bidder with the highest bid $v$, and pays the price of the second highest bid, $v'$. 
Bidders with assigned items are, therefore, able to call \algorithm{Withdraw} to withdraw $(v-v')$ coins.
\Bidders with assigned items can now contact sellers and prove their identity by signing any message using the private key used to participate in the auction (see \Cref{sec:commit}).

\begin{figure}[t]
\includegraphics[width=\linewidth]{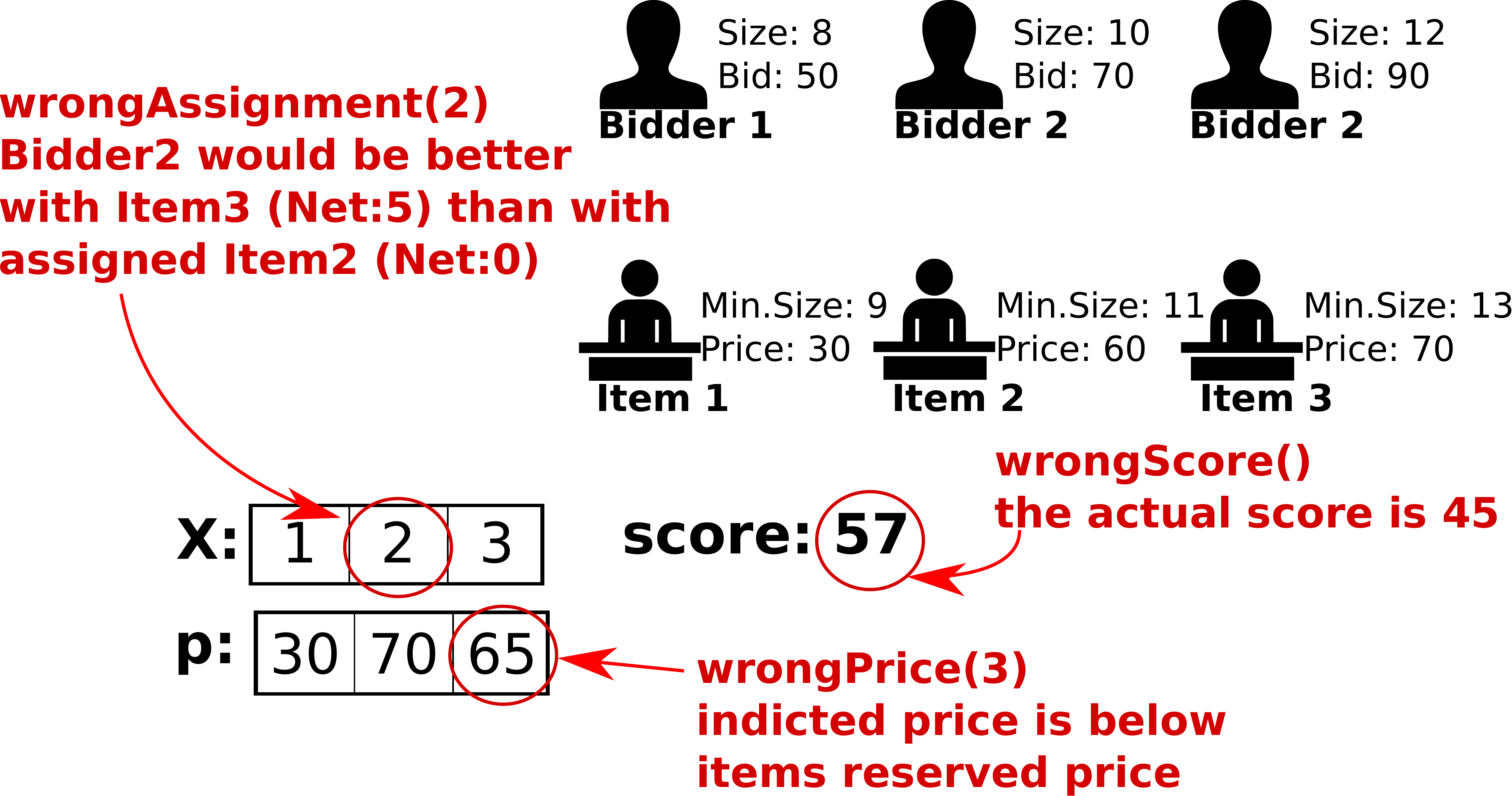}
\caption{An incorrect solution with different submitted proofs of misbehaviour.
}
\label{fig:proofs}
\end{figure}

\para{Misbehaviour example}
\Cref{fig:proofs} presents an example of an incorrect solution submitted to the smart contract for an auction with 3 bidders and 3 items.
The assignment \textbf{X} is not an equilibrium.
Net valuation (the difference between the bid and the price) of Bidder~2 for its currently allocated Item~2 equals to 0 and is lower than for Item~3 (net valuation of 5).
A valid proof of misbehaviour would point to the user, allocated Item~2 and alternative Item~3, allowing the smart contract to calculate and compare both valuations efficiently.
Furthermore, the price (65) of Item~3 is incorrect as it is below the reservation price specified by the seller (70).
A valid proof of misbehaviour simply points to the invalid price allowing the smart contract to verify it against the reservation price and the bid expressed by the assigned bidder.
Finally, the declared score is incorrect, as the sum of net evaluations is 45 ($20+0+25$). 

%% file: sections/security.tex
\section{Protocol Analysis and Discussion}
\label{sec:analysis}

We argue that \sysname achieves the design goals described in \Cref{sec:goals}.

\para{Auction's economic properties}
An auction satisfies all such properties only under the condition of \textit{price-taker participants}~\cite{myerson1983efficient}, \ie both \bidders and \sellers have no impact on bids expressed by other users.
In the multi-item auctions we consider, it has been proven that Vickrey auctions possesses all these desired attributes based on the assumption of ``sealed bids'', where neither \bidders nor \sellers have information about the state of the auction~\cite{vickrey1961counterspeculation}.
\sysname through its privacy properties provides a technical implementation of this ``sealed bids'' assumption, and prevents price manipulations.

Our proofs of misbehaviours cover the full spectrum of potential misbehaviours in terms of proposed solutions.
\algorithm{wrongAssignment} and \algorithm{wrongPrice} limit a user to submit only valid equilibrium solutions respecting minimum prices.
That is, the mechanism cannot be unfair to either a participating bidder, who is happy with its assigned items at the given price, or a seller, whose item can only be offered at more than its reservation price. 
Given the ability to filter solutions that do not fall into the valid equilibrium category, the involved score-based system elicits the VCG equilibrium by allowing anyone to submit the optimal solution with the highest score.
At the same time, an incorrect score can be contested using \algorithm{wrongScore}.
In other words, in \sysname the dedicated node cannot exploit the mechanism by submitting an invalid and/or unfair solution while it is incentivised to submit the VCG equilibrium.

\para{Correctness} \sysname is implemented as a smart contract; its correct execution can be verified by any third party, taking advantage of the \emph{Public Auditability} of the blockchain.

\Bidders are bound to their bids as they are first required to commit to their bid, and then to open the commitment by unblinding the signature providing \emph{Bids binding}.
\Bidders cannot open the commitment to another value than the previously committed, as this would require forging the underlying signature.
The users commit to their bid during the commit phase, thus they cannot modify it after observing the bids of other users. 

No single \authority can issue a signature and steal all the coins in the smart contract---the threshold property of the signature implies that adversaries need to corrupt an arbitrarily large set of authorities for this attack to be possible.
\Bidders cannot participate in the auction (\Cref{sec:preparation}) without paying a deposit to receive a valid signature.
The timer forces the \bidders to commit to their bid before revealing it, preventing any third party from seeing other \bidder's bid before committing to a value.
\Bidders dropping out after committing a bid (and never revealing it) are financially penalized as they cannot withdraw their coins.
Once issued, signatures can only be used for the intended auction since they embedded an identifier $\mathit{id}$ unique to this auction.
As a result, users that do not follow the protocol (\ie first commit, and then reveal) lose the associated coins, guaranteeing \emph{Fairness}.

\para{Privacy} Upon submission of item descriptions, sellers provide a commitment to their minimum prices instead of the actual values. The minimum price is revealed only during the reveal phase when no additional bids are accepted (\emph{Hidden Minimum Price}). 

\sysname takes advantage of the unlinkability property of the underlying blind signature to break the link between the commit and reveal phase, provided that \bidders use fresh addresses. As a result, \bidders can submit bids on auctions without revealing their identity enabling \emph{\Bidders' privacy}. 

Bids are kept private until the timer expires; the blindness property of the underlying signature scheme implies that no information about the bid is revealed while committing to a bid (\emph{Bids privacy}).

\para{Deployment} During the preparation phase (\Cref{sec:preparation}), \Bidders only interact with a subset of the \authorities; during the execution phase (\Cref{sec:execution}), they only interact with the smart contract; during the execution phase, \bidders only interact with the \sellers or with the smart contract to withdraw their coins (\emph{Non-interactivity}).

The decentralized nature of the blockchain makes the \sysname smart contract resilient to censorship and guarantees \emph{Openness}.
Furthermore, a small subset of \authorities cannot block the issuance of blind signatures---the service is guaranteed to be available as long as at least a threshold number of authorities are available.

\sysname does not introduce any single trusted third party; the \sysname contract is executed on a decentralized smart contract platform, and allows threshold issuance by \emph{Distributed authorities}.

\sysname offloads the most CPU-intensive computation to off-chain nodes, decreasing the operations performed on the Smart Contract.
The cost of almost all Smart Contract methods is not influenced by increasing the number of items and/or bidders.
The only two exceptions consist of \algorithm{submitSolution} and \algorithm{wrongScore}.
The former requires increased on-chain storage when the number of participants increases.
However, this cost can be covered from a small commission on system execution.
The latter is invoked only when an incorrect solution is proposed and its cost is completely covered by the misbehaving node (\emph{Scalability}).
The off-chain solution calculation does not introduce any overhead compared to the original and optimal version of the algorithm.
Finally, the verification done by users requires orders of magnitude lower calculation time than calculating the solution itself.
We provide an extensive evaluation of all the \sysname components in \Cref{sec:evaluation}.

\para{Extension to other types of auctions}
\sysname can be easily extended to support both single-item and combinatorial auctions. 
Single-item auctions can be seen as a special case of multi-item auctions (with one item only) and does not require any further modifications. 
The main difference between multi-item and combinatorial auctions lies in the computations' complexity.
While it is possible to deterministically derive an optimal solution when using multi-item auction, it has been shown that combinatorial auctions can be modelled as the set packing problem, meaning that they are NP-hard and there is no polynomial-time algorithm for finding the optimal allocation~\cite{lehmann2006winner}.
However, \sysname does not perform auction assignment computations on-chain. 
The score used could be computed in a similar way to our current design, but with a different set of constraints.
Submitting solutions to a combinatorial auction can be seen as a proof of useful work, where better solutions replace those with lower scores.
Most of the changes are applied to the off-chain component, to implement algorithms computing assignments and verifying solutions submitted by others. 

\para{Non-rational attackers}
Non-rational bidders can spend resources to influence the market, \ie by buying the majority of available storage when they do not need it. Such behavior would temporarily inflate the prices for the scarce resource following the rules of supply and demand. However, \sysname guarantees that the bids are binding (the attacker will pay for the acquired storage) and all bids are sealed/private (the attacker has no way of precisely targeting specific prices/bidders). Non-rational behaviors cannot break these guarantees, and \sysname provides the same level of resilience to market-based attacks as VDA it employs (\ie bidding strategies are intrinsic to open-market environments). Finally, higher prices paid to workers would encourage additional sellers to join the network, increase the supply and restore the price equilibrium in the long run.

%% file: sections/evaluation.tex
\section{Implementation \& Evaluation}
\label{sec:evaluation}
We evaluate the \sysname prototype on a desktop device for the user side and Ethereum for the Smart Contract platform, and compare it with an implementation of Verifiable Sealed-bid Auction (VSA)~\cite{galal2018verifiable}. VSA provides similar auditability guarantees, follows a similar to ours approach running heavy computations off-chain, but supports only single-item auctions\footnote{This section still evaluates a multi-item version of our system}. We are not aware of any system providing similar guarantees as ours, including privacy, for multi-item auctions. Note that general, verifiable computation platforms require validators to repeat the computations~\cite{kalodner2018arbitrum}, do not provide sufficient privacy protection~\cite{teutsch2016truebit} or expierience multiple orders of magnitude higher complexity~\cite{kosba2016hawk}

We implement \sysname's off-chain component in Python and the Smart Contract part in Solidity~\cite{pastrami}. We use Web3py~\cite{web3py} for communication between both modules. The implementation consists of 1000 Python LoC and 410 Solidity LoC including a simple command-line user interface. 
We rely on threshold BLS signatures~\cite{bgls} which are blinded using a construction introduced by Boldyreva~\cite{boldyreva2003threshold}. 
This construction provides unlinkability between the signing and verification process and preserves users' anonymity.
We use a slightly modified version of the original blind BLS signature of Boldyreva in order to make it compatible with type-3 pairings, which are efficient~\cite{galbraith2008pairings} and supported by \ethereum as pre-compiled contracts (for optimal Ate pairing checks) on the elliptic curve alt\_bn128~\cite{eip197}.


\para{Setup}
We deploy the \sysname Smart Contract on the Ethereum Ropsten Testnet.
The application listens for events generated on-chain.
It can automatically compute a solution after the reveal phase, or verify solutions submitted by others.
All off-chain computations are performed on a Dell Latitude 5590 laptop with an Intel Core i7-8650U CPU and 16~GB of RAM. 

The VSA~\cite{galal2018verifiable} implementation uses two Ethereum Smart Contracts and a client application written in C\#~\cite{auctioneer}. 

\begin{figure}[t]
\begin{minipage}[t]{.4\linewidth}
  \includegraphics[width=1\linewidth]{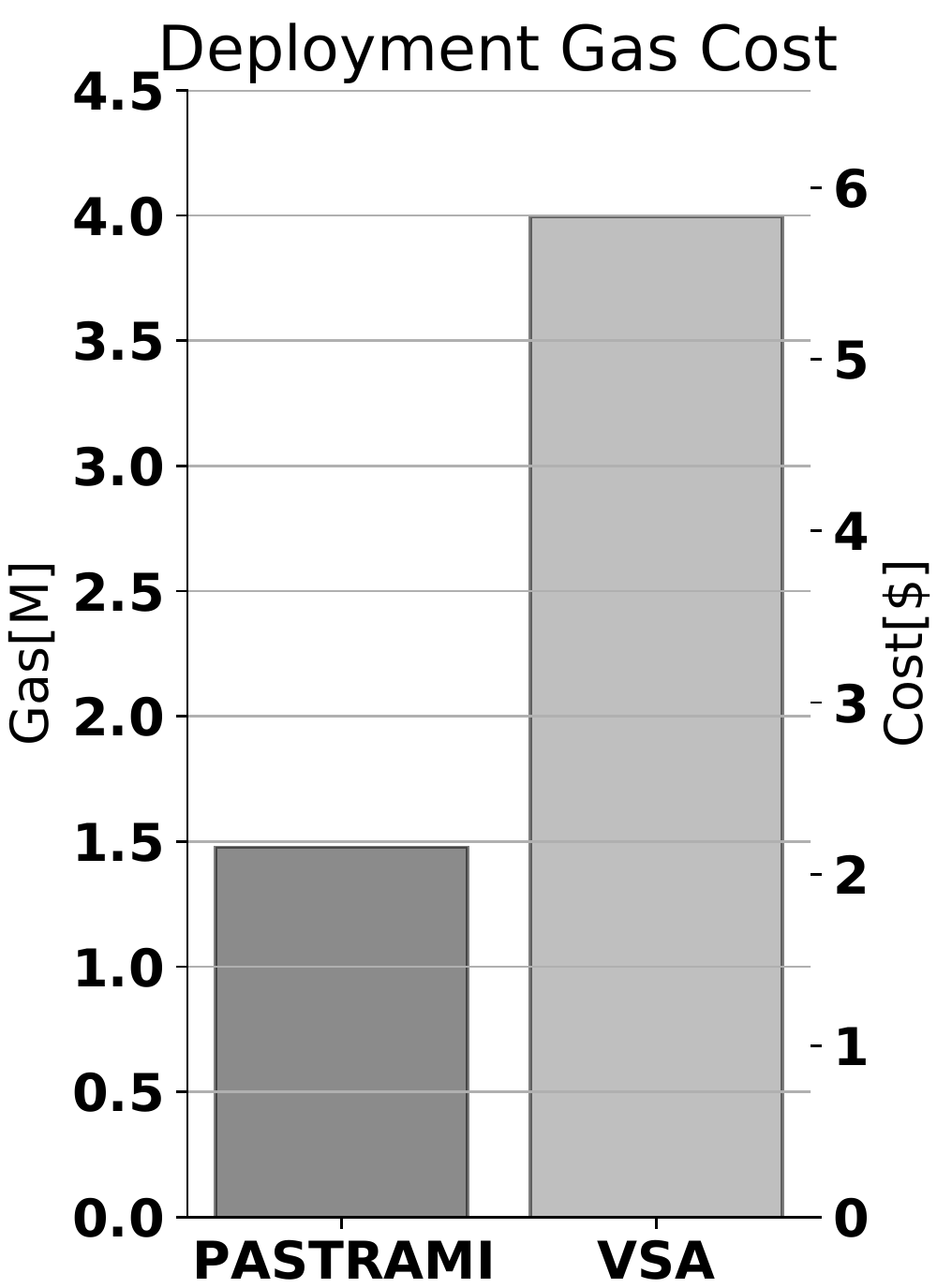}
  \caption{Deployment cost.}
  \label{fig:comparison1}
\end{minipage}%
\begin{minipage}[t]{.6\linewidth}
  \centering
  \includegraphics[width=1\linewidth]{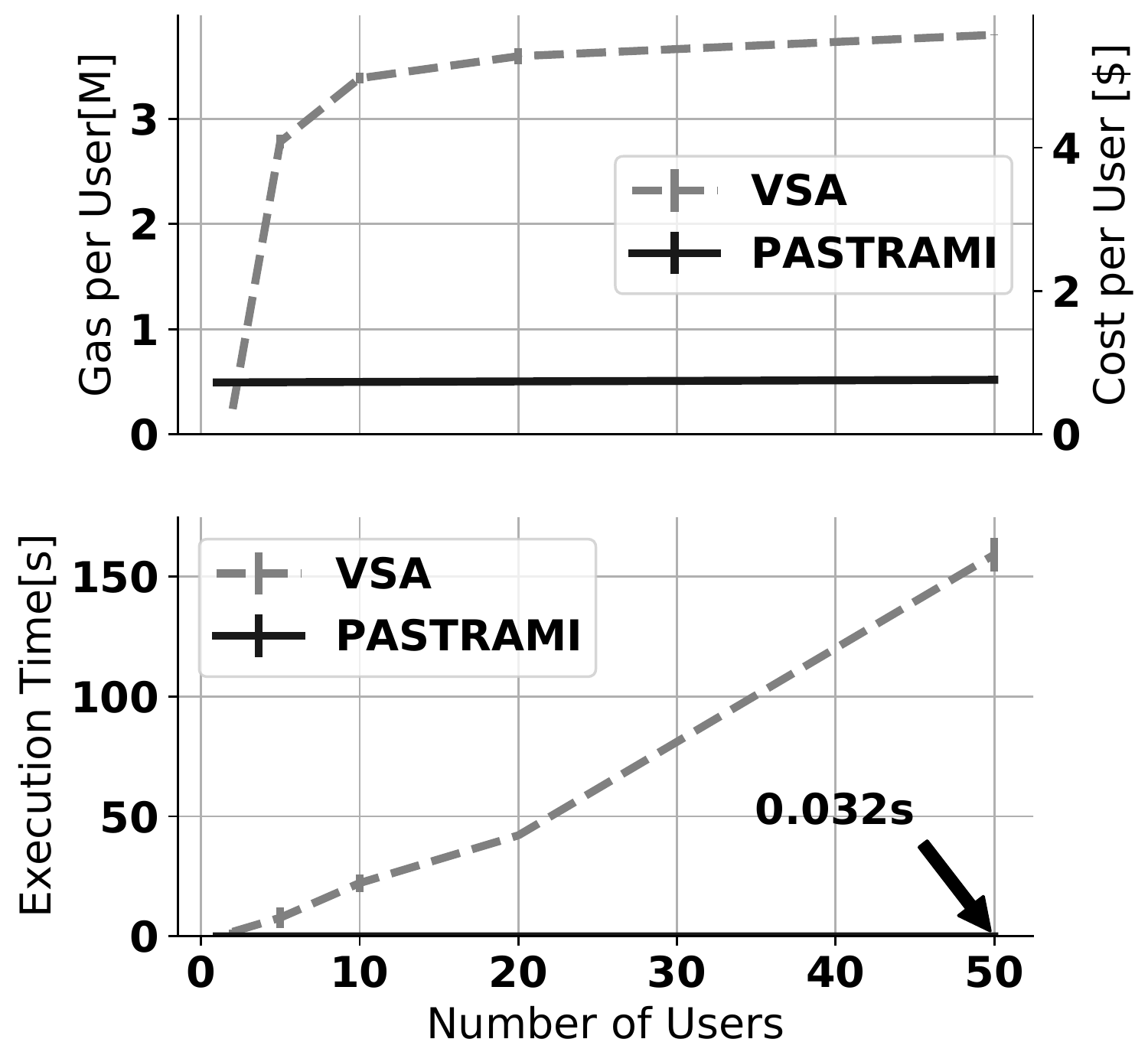}
  \caption{Execution cost.}
  \label{fig:comparison2}
 \end{minipage}%
\end{figure}

\para{Comparison with VSA}
We start by investigating the cost of deploying both platforms on Ethereum (\Cref{fig:comparison1}).
Due to much lower complexity and smaller contract size, \sysname incurs a 2.5 times lower deployment cost than VSA. 

We continue by performing single-item auctions with increasing number of users, as VSA only supports this type of auctions.
We measure the gas cost for each bidder (\Cref{fig:comparison2} top).
The \sysname Smart Contract performs only a simple set of operations, related to submitting and revealing bids.
It keeps the cost per user low and almost constant when the number of participants increases.
In contrast, VSA executes the auction algorithm on-chain, resulting in much higher cost, with a per-bidder costs that raises with the number of bidders. 

Finally, we compare the execution time of the off-chain components for all the users and the node calculating solution in \sysname (\Cref{fig:comparison2} bottom).
Even though \sysname needs to calculate an assignment off-chain, the required operations are much simpler than the costly zero-knowledge proofs used by VSA. 
It results in much lower computational load and much lower impact of an increasing number of users.

\para{Smart Contract Evaluation}
We investigate the cost of invoking each function of the \sysname Smart Contract on Ethereum.
We use Remix~\cite{remix} to calculate this cost in gas and ETH Gast Station~\cite{gas_station} to convert it to ETH and to an indicative amount in USD\footnote{Measured on November 30th 2019.}. 
For the conversion, we use Ethereum Slow mode which increases confirmation time, but decreases monetary cost of all the operations.
\Cref{tab:evaluation} presents the results.
We split the number into fixed part (not influenced by the number of bidders/items) and dynamic part expressed as an additional cost per item/bidder.
For simplicity, we assume the same number of bidder and items participating in the auction.
The presented cost includes both the transaction and the execution cost.

The \algorithm{Commit} algorithm is cheap as it only performs basic checks and emits an event asking the authorities to issue a blind signature to the user; our implementation assumes the signature issuance happens off-chain.
The \algorithm{Reveal} algorithm is more expensive as it performs a signature verification, involving elliptic curve pairing checks. 
The analogical algorithms for submitting and revealing item description involves only simple commitments and much lower monetary cost.

Submitting a solution with \algorithm{submitSolution} consists only of storing the solution on-chain.
This involves a small static cost and small cost per item for the storage space. 
\algorithm{wrongAssignment} and \algorithm{wrongPrice} use indexes passed as input parameters and perform only simple mathematical operations (comparison and subtractions).
Finally, \algorithm{wrongScore} requires iterating through the assignment and re-calculating the score in order to verify whether the declared score is correct.
The minimum collateral submitted by the dedicated node should be greater than the cost of the most expensive method being part of the proofs of misbehaviour. \sysname automatically computes this value based on information from \Cref{tab:evaluation} and the size of the auction and does not accept solutions with lower collateral.  

%

\begin{table}[t]
 
\newcolumntype{C}{>{\raggedright\let\newline\\\arraybackslash\hspace{0pt}}m{0.3\linewidth} }
\newcolumntype{D}{>{\raggedright\arraybackslash} m{0.25\linewidth} }

 \centering
 \begin{tabular}{C D D}
     \toprule
     \bf Operation & \bf Gas & \bf USD\\
     \midrule
     \textsf{Commit} &26,590  & 0.15\\ 
     \textsf{Reveal} &364,456  & 2.06 \\ 
     \textsf{submitItem} &43,556  & 0.3 \\ 
     \textsf{RevealMinPrice} & 52,378  & 0.361 \\ 
     \textsf{submitSolution} &5,068 + 408* &0.155 + 0.003*\\ 
     \textsf{wrongAssignment} & 45,572 & 0.282 \\ 
     \textsf{wrongScore} &18,048 + 6,494* & 0.112 + 0.04* \\ 
     \textsf{wrongPrice} & 35,714 & 0.221\\ 
   
     \bottomrule
 \end{tabular}
 \caption{Cost of invoking \sysname functions. *per bidder/item}
\label{tab:evaluation}
\end{table}

\para{Filecoin deployment}
As an example of the use of auctions in a decentralised utility platform, we extract data about storage nodes from the Filecoin testnet~\cite{benet2014ipfs, benet2018proof}.
At the core of Filecoin lies a \emph{Storage Market} that acts as an exchange point where clients and miners can advertise their requests and offer storage space.
Currently, Filecoin maintains an on-chain order book storing a public list of active storage miner offers and client orders.
It rely on a direct, peer-to-peer off-chain assignment between users and storage providers.
Once two parties reach an agreement, they upload a signed deal to the blockchain.

We query the Filecoin network for all the miners offering storage together with their features.
With the current API, we were able to get the current storage price per GB and storage duration, but the available storage capacity is currently unavailable~\cite{filecoin_storage_issue}.

We perform multi-item Vickrey-Dutch auctions using different numbers of bidders.
We convert Filecoin prices to USD per GB per month.
The bids are derived from the Google Cloud Storage price~\cite{google_storage_cmp} and randomized storage duration such that bidders willing to rent storage space for longer pay a lower price per month.

\begin{figure}[t]
\includegraphics[width=\linewidth]{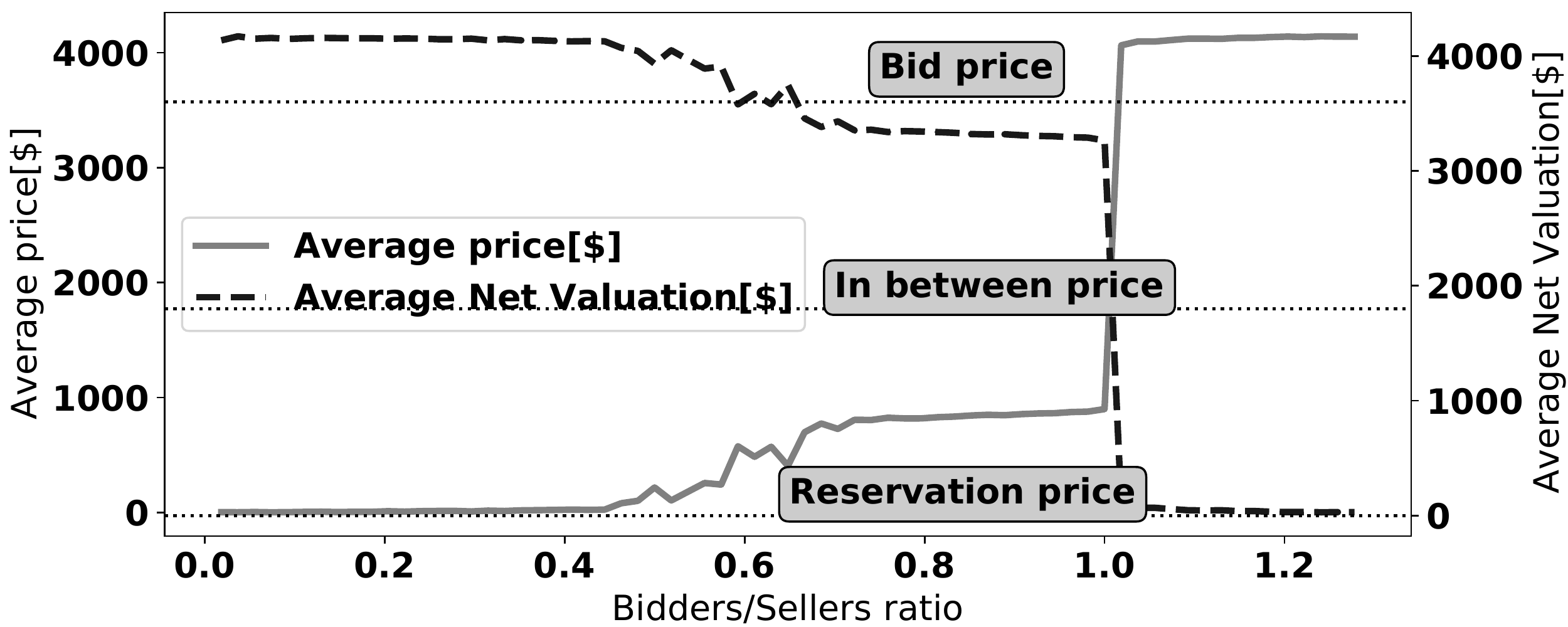}
\caption{Prices and net valuation derived by \sysname.}
\label{fig:filecoin_auction}
\end{figure}

\Cref{fig:filecoin_auction} presents the evaluation of average price and average net valuation in USD.
Dotted lines indicate simpler mechanisms where bidders manually contact advertised workers and agree to pay reservation price, valuation price or value in between.
We observe that the auction algorithm is able to adapt to changes in the supply/demand ratio and derive the most optimal prices.
When the number of bidders exceeds the number of items, the price raises to the average valuation and does not increase further, as we assume that at this point, users will switch to cloud storage providers. 

\begin{figure}[t]
\includegraphics[width=\linewidth]{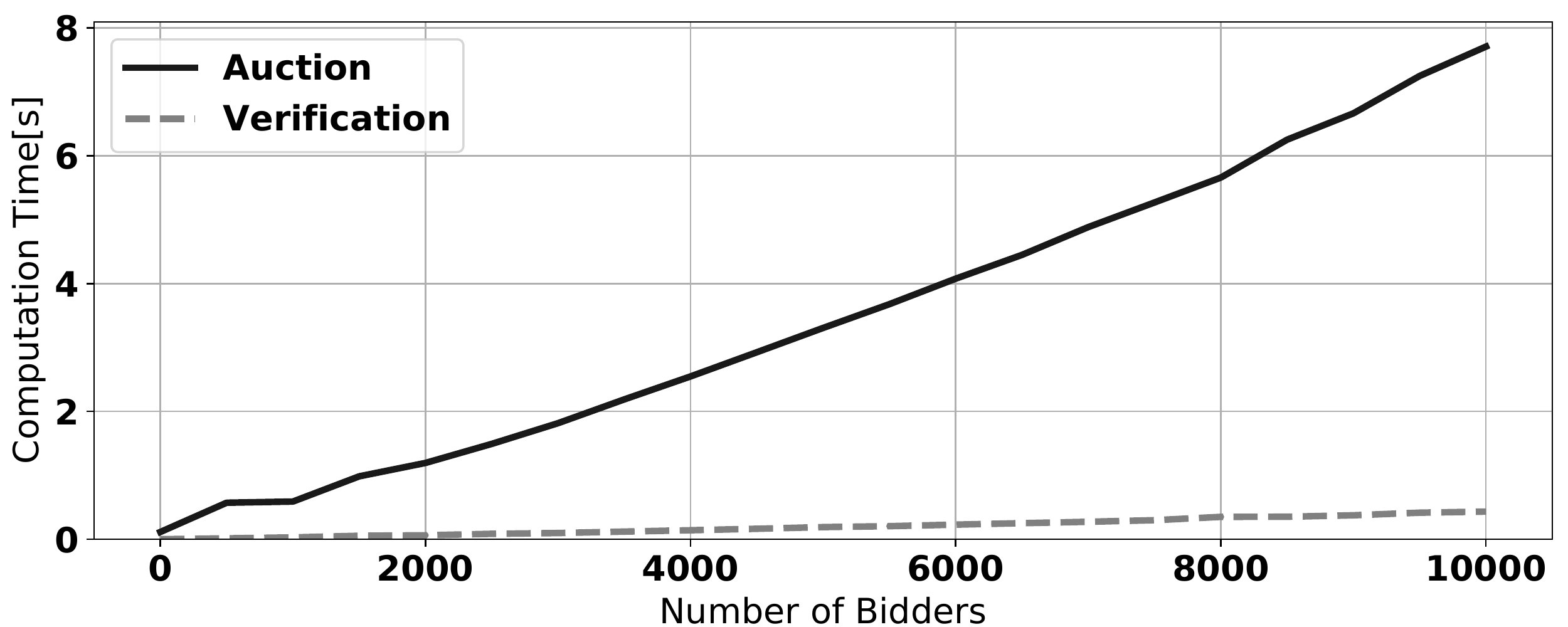}
\caption{Execution time of auction and verification algorithms.}
\label{fig:filecoin_cpu}
\end{figure}

Finally, we measure the time required to calculate a solution and verify it with increasing number of users (\Cref{fig:filecoin_cpu}).
Even for 10,000 users, \sysname is able to derive an optimal solution in less than 8s.
Furthermore, the verification process proves to be efficient and completes within 500ms even for the largest auction. 

%% file: sections/related.tex
\section{Related Work}\label{sec:related}
\newcommand{\yestick}{{\color{green}{\cmark}}}
\newcommand{\notick}{{\color{red}{\xmark}}}

\begin{table*}[t]
\centering
\resizebox{\textwidth}{!}{
\begin{tabular}{  l  cccccc}
\toprule
\bf System & \bf Bids Privacy & \bf Bidders' Privacy & \bf Non-Interactivity & \bf Distributed Authority & \bf Trusted Hardware & \bf Public Auditability\\ 
\midrule    

ShadowEth~\cite{yuan2018shadoweth} & \yestick & \notick  & \yestick & \H & Intel SGX~\cite{costan2016intel} & \yestick\\
Hawk~\cite{kosba2016hawk} & \yestick & \notick&  \yestick & \M & None  & \yestick\\
Strain~\cite{blass2018strain} & \yestick & \notick & \notick & \L & None  & \yestick\\
Galal~\etal~\cite{galal2018succinctly} & \yestick & \notick & \notick & \L & None  & \yestick\\
Bogetoft~\etal~\cite{bogetoft2009secure} & \yestick & \notick & \yestick & \H & None  & \notick\\
Filecoin~\cite{filecoin} & \notick & \notick & \notick & \H & None & \yestick\\
Galal \etal~\cite{galal2018verifiable} & \yestick & \notick & \yestick & \L & None & \yestick \\
\rowcolor{verylightgray}
\textbf{\sysname} & \yestick & \yestick & \yestick & \H & None & \yestick \\
\bottomrule
\end{tabular}}

\caption{\footnotesize Comparison of properties achieved by related systems. The decentralisation property reads as follows; \L:~relies on a trusted third party, \\ \M:~relies on a trusted third party for only one (or some) of the properties described in \Cref{sec:goals}, \H:~does not rely on any trusted third party.} 
\label{tab:related}
\end{table*}

We cover related work on private computation over blockchains, auction platforms, and decentralised cloud platforms.
We summarize the most related solutions and their security features in \Cref{tab:related}.

Several frameworks aim at hiding private data submitted to a public ledger.
Hawk~\cite{kosba2016hawk} divides Smart Contracts into public and private parts and secure private input using zero-knowledge proofs, but requires a centralized trusted manager to operate.
ShadowEth~\cite{yuan2018shadoweth} allows processing confidential Smart Contract data using Trusted Execution Environments (TEE).
However, such a scheme requires users to trust the hardware vendor and can expose the system to TEE's vulnerabilities~\cite{wang2017leaky,chen2019sgxpectre,van2018foreshadow}.

Blass and Kerschbaum~\cite{blass2018strain} propose Strain, a system that preserves sealed-bid auctions privacy against malicious participants.
Strain uses a two-party comparison protocol based on Secure Multi-Party Computation (MPC), but has a flaw that reveals the order of bids.
Furthermore, running protocols involving MPC on blockchain is not efficient due to the extensive computations and the number of rounds involved.
Galal and Youssef~\cite{galal2018verifiable} present a protocol ensuring public verifiability, privacy of bids, and fairness.
However, the solution scales badly with the number of bidders and relies on random number retrieved from blockchains that are not proven to be secure.
This scheme was later improved using zk-SNARKS~\cite{galal2018succinctly}, but it still relies on a centralized party for zero-knowledge proofs and does not protect bidders' identities.
An alternative approach was proposed by Bogetoft \etal~\cite{bogetoft2009secure}.
Their system uses MPC to perform auctions on encrypted bids.
However, such a scheme reveals the final assignment between bidders and items and does not provide transparency. 

Filecoin~\cite{filecoin} does not implement an automated system assigning clients to storage nodes.
Users are required to chose storage nodes on a one-to-one basis and offers are publicly posted on the blockchain.
Other industrial systems such as Golem~\cite{golem}, iExec~\cite{iexec} or SONM~\cite{sonm} either do not specify their requester---worker assignment technique or rely on similar, non-transparent solutions.
All those platform could use \sysname to increase their level of security and automatically determine optimal price for services.

%% file: sections/conclusion.tex
\section{Conclusion}\label{sec:conclusion}


We presented \sysname, a system for determining an assignment between providers and users and deriving optimal prices in a decentralised environment, while preserving the privacy and accountability of participants.

\sysname enables a transparent use of Vickey-Dutch multi-item auctions to derive prices, assignments and maximise social wellness.
It secures users bids as well as the identity of the bidders using blind signatures. 
In contrast with previous work, \sysname does not rely on a trusted third party to issue signatures, but rather on a set of entities that can be freely chosen by users.
These distributed authorities issue only partial signatures that are merged locally by each users, protecting the system from a subset of malicious authorities. 

To enable scalability, heavy computations are performed off-chain and submitted to a smart contract.
An assignment for an auction can be challenged by proofs of misbehaviour submitted by users to this smart contract.
We have shown how \sysname can be deployed on Ethereum blockchain and adapted to shared economy systems with the example of Filecoin, and our evaluation shows its reduced costs compared to a system with similar auditability guarantees.

%% file: appendices/blind_bls.tex


\section{Blind BLS signatures}
\label{sec:blind-bls}

We recall the cryptographic construction of the blind BLS signature~\cite{boldyreva2003threshold} used in our implementation.
For the sake of simplicity, we describe below a key generation algorithm \algorithm{BS.TTPKeyGen} as executed by a trusted third party; this protocol can however be executed in a distributed way as illustrated by Gennaro~\etal~\cite{gennaro1999secure} under a synchrony assumption, and as illustrated by Kate~\etal~\cite{cryptoeprint:2012:377} under a weak synchrony assumption.

\begin{description}[leftmargin=1em, labelindent=0em]
\setlength\itemsep{0.5em}
\item[\definition{BS.Setup($1^\lambda$)}{$\mathit{params}$}] Choose a bilinear group \\ $(\mathbb{G}_1,\mathbb{G}_2,\mathbb{G}_T)$ with order $p$, where $p$ is an $\lambda$-bit prime number. Let $g_1$ be a generator of $\mathbb{G}_1$, and $g_2$ a generator of $\mathbb{G}_2$. The system parameters are $\mathit{params}=(\mathbb{G}_1, \mathbb{G}_2, \mathbb{G}_T, p, g_1, g_2)$. 

\item[\definition{BS.TTPKeyGen($params, t, n$)}{$x, y$}] Pick a polynomial $v$ of degree $t-1$ with coefficients in $\mathbb{F}_p$, and set $(x, y) = (v(0), g_2^{v(0)})$. Issue to each authority $i \in [1, \dots, n]$ a secret key $x_i = v(i)$, and publish their verification key $y_i = g_2^{x_i}$.

\item[\definition{BS.PrepareBlindSign($m$)}{$\tilde{h}$}] pick a random $r \leftarrow \mathbb{F}$; output $\tilde{h}=\hashtopoint(m)^r \in \mathbb{G}_1$.

\item[\definition{BS.BlindSign($x_i, \tilde{h}$)}{$\tilde{\sigma}_i$}] output $\tilde{\sigma}_i = \tilde{h}^{x_i}$.

\item[\definition{BS.Unblind($r, \tilde{\sigma}$)}{$\sigma$}] output $\sigma = \tilde{\sigma}^{(-r)}$.

\item[\definition{BS.AggSig($\sigma_1, \dots, \sigma_t$)}{$\sigma$}] Output $\sigma = \prod^t_{i=1} \sigma_i^{l_i})$, where $l$ is the Lagrange coefficient:
\begin{equation}\nonumber
l_i = \left[\prod^t_{j=1, j\neq i} (0-j)\right] \left[\prod^t_{j=1, j\neq i} (i-j)\right]^{-1} \;{\rm mod}\;
\end{equation}

\item[\definition{BS.Verify($y, m, \sigma$)}{$\mathit{true}/\mathit{false}$}] compute $h=\hashtopoint(m)$; output $\mathit{true}$ if $e(h, y) = e(\sigma, g_2)$; otherwise output $\mathit{false}$.
\end{description}

%% file: auctions.bbl

\begin{thebibliography}{56}


\ifx \showCODEN    \undefined \def \showCODEN     #1{\unskip}     \fi
\ifx \showDOI      \undefined \def \showDOI       #1{#1}\fi
\ifx \showISBNx    \undefined \def \showISBNx     #1{\unskip}     \fi
\ifx \showISBNxiii \undefined \def \showISBNxiii  #1{\unskip}     \fi
\ifx \showISSN     \undefined \def \showISSN      #1{\unskip}     \fi
\ifx \showLCCN     \undefined \def \showLCCN      #1{\unskip}     \fi
\ifx \shownote     \undefined \def \shownote      #1{#1}          \fi
\ifx \showarticletitle \undefined \def \showarticletitle #1{#1}   \fi
\ifx \showURL      \undefined \def \showURL       {\relax}        \fi
\providecommand\bibfield[2]{#2}
\providecommand\bibinfo[2]{#2}
\providecommand\natexlab[1]{#1}
\providecommand\showeprint[2][]{arXiv:#2}

\bibitem[\protect\citeauthoryear{??}{iex}{2018}]%
        {iexec}
 \bibinfo{year}{2018}\natexlab{}.
\newblock \bibinfo{title}{{iExec} Whitepaper}.
\newblock
  \bibinfo{howpublished}{\url{https://iex.ec/whitepaper/iExec-WPv3.0-English.pdf}}.
\newblock


\bibitem[\protect\citeauthoryear{??}{son}{2018}]%
        {sonm}
 \bibinfo{year}{2018}\natexlab{}.
\newblock \bibinfo{title}{{SONM}: Decentralized Fog Computing Platform}.
\newblock \bibinfo{howpublished}{\url{https://sonm.com}}.
\newblock


\bibitem[\protect\citeauthoryear{Andrychowicz, Dziembowski, Malinowski, and
  Mazurek}{Andrychowicz et~al\mbox{.}}{2014}]%
        {andrychowicz2014secure}
\bibfield{author}{\bibinfo{person}{Marcin Andrychowicz},
  \bibinfo{person}{Stefan Dziembowski}, \bibinfo{person}{Daniel Malinowski},
  {and} \bibinfo{person}{Lukasz Mazurek}.} \bibinfo{year}{2014}\natexlab{}.
\newblock \showarticletitle{Secure multiparty computations on bitcoin}. In
  \bibinfo{booktitle}{\emph{2014 IEEE Symposium on Security and Privacy}}.
  IEEE, \bibinfo{pages}{443--458}.
\newblock


\bibitem[\protect\citeauthoryear{Benet}{Benet}{2014}]%
        {benet2014ipfs}
\bibfield{author}{\bibinfo{person}{Juan Benet}.}
  \bibinfo{year}{2014}\natexlab{}.
\newblock \showarticletitle{Ipfs-content addressed, versioned, p2p file
  system}.
\newblock \bibinfo{journal}{\emph{arXiv preprint arXiv:1407.3561}}
  (\bibinfo{year}{2014}).
\newblock


\bibitem[\protect\citeauthoryear{Benet, Dalrymple, and Greco}{Benet
  et~al\mbox{.}}{2018}]%
        {benet2018proof}
\bibfield{author}{\bibinfo{person}{Juan Benet}, \bibinfo{person}{David
  Dalrymple}, {and} \bibinfo{person}{Nicola Greco}.}
  \bibinfo{year}{2018}\natexlab{}.
\newblock \bibinfo{booktitle}{\emph{Proof of replication}}.
\newblock \bibinfo{type}{{T}echnical {R}eport}. \bibinfo{institution}{Technical
  report, Protocol Labs, July 27, 2017. https://filecoin.
  io/proof-of-replication. pdf. Accessed June}.
\newblock


\bibitem[\protect\citeauthoryear{Bissias, Ozisik, Levine, and
  Liberatore}{Bissias et~al\mbox{.}}{2014}]%
        {xim}
\bibfield{author}{\bibinfo{person}{George Bissias}, \bibinfo{person}{A.~Pinar
  Ozisik}, \bibinfo{person}{Brian~N. Levine}, {and} \bibinfo{person}{Marc
  Liberatore}.} \bibinfo{year}{2014}\natexlab{}.
\newblock \showarticletitle{Sybil-Resistant Mixing for Bitcoin}. In
  \bibinfo{booktitle}{\emph{Workshop on Privacy in the Electronic Society}}
  \emph{(\bibinfo{series}{WPES '14})}. \bibinfo{publisher}{ACM},
  \bibinfo{pages}{149--158}.
\newblock
\showISBNx{978-1-4503-3148-7}


\bibitem[\protect\citeauthoryear{Blass and Kerschbaum}{Blass and
  Kerschbaum}{2018}]%
        {blass2018strain}
\bibfield{author}{\bibinfo{person}{Erik-Oliver Blass} {and}
  \bibinfo{person}{Florian Kerschbaum}.} \bibinfo{year}{2018}\natexlab{}.
\newblock \showarticletitle{Strain: A secure auction for blockchains}. In
  \bibinfo{booktitle}{\emph{European Symposium on Research in Computer
  Security}}. Springer, \bibinfo{pages}{87--110}.
\newblock


\bibitem[\protect\citeauthoryear{Bogetoft, Christensen, Damg{\aa}rd, Geisler,
  Jakobsen, Kr{\o}igaard, Nielsen, Nielsen, Nielsen, Pagter,
  et~al\mbox{.}}{Bogetoft et~al\mbox{.}}{2009}]%
        {bogetoft2009secure}
\bibfield{author}{\bibinfo{person}{Peter Bogetoft}, \bibinfo{person}{Dan~Lund
  Christensen}, \bibinfo{person}{Ivan Damg{\aa}rd}, \bibinfo{person}{Martin
  Geisler}, \bibinfo{person}{Thomas Jakobsen}, \bibinfo{person}{Mikkel
  Kr{\o}igaard}, \bibinfo{person}{Janus~Dam Nielsen},
  \bibinfo{person}{Jesper~Buus Nielsen}, \bibinfo{person}{Kurt Nielsen},
  \bibinfo{person}{Jakob Pagter}, {et~al\mbox{.}}}
  \bibinfo{year}{2009}\natexlab{}.
\newblock \showarticletitle{Secure multiparty computation goes live}. In
  \bibinfo{booktitle}{\emph{Intl. Conf. on Financial Cryptography and Data
  Security}}. Springer, \bibinfo{pages}{325--343}.
\newblock


\bibitem[\protect\citeauthoryear{Boldyreva}{Boldyreva}{2003}]%
        {boldyreva2003threshold}
\bibfield{author}{\bibinfo{person}{Alexandra Boldyreva}.}
  \bibinfo{year}{2003}\natexlab{}.
\newblock \showarticletitle{Threshold signatures, multisignatures and blind
  signatures based on the gap-Diffie-Hellman-group signature scheme}. In
  \bibinfo{booktitle}{\emph{International Workshop on Public Key
  Cryptography}}. Springer, \bibinfo{pages}{31--46}.
\newblock


\bibitem[\protect\citeauthoryear{Boneh, Gentry, Lynn, and Shacham}{Boneh
  et~al\mbox{.}}{2003}]%
        {bgls}
\bibfield{author}{\bibinfo{person}{Dan Boneh}, \bibinfo{person}{Craig Gentry},
  \bibinfo{person}{Ben Lynn}, {and} \bibinfo{person}{Hovav Shacham}.}
  \bibinfo{year}{2003}\natexlab{}.
\newblock \showarticletitle{Aggregate and verifiably encrypted signatures from
  bilinear maps}. In \bibinfo{booktitle}{\emph{Eurocrypt}},
  Vol.~\bibinfo{volume}{2656}. Springer, \bibinfo{pages}{416--432}.
\newblock


\bibitem[\protect\citeauthoryear{Boneh, Lynn, and Shacham}{Boneh
  et~al\mbox{.}}{2001}]%
        {bls}
\bibfield{author}{\bibinfo{person}{Dan Boneh}, \bibinfo{person}{Ben Lynn},
  {and} \bibinfo{person}{Hovav Shacham}.} \bibinfo{year}{2001}\natexlab{}.
\newblock \showarticletitle{Short signatures from the Weil pairing}.
\newblock \bibinfo{journal}{\emph{ASIACRYPT}} (\bibinfo{year}{2001}),
  \bibinfo{pages}{514--532}.
\newblock


\bibitem[\protect\citeauthoryear{Bonneau, Narayanan, Miller, Clark, Kroll, and
  Felten}{Bonneau et~al\mbox{.}}{2014}]%
        {mixcoin}
\bibfield{author}{\bibinfo{person}{Joseph Bonneau}, \bibinfo{person}{Arvind
  Narayanan}, \bibinfo{person}{Andrew Miller}, \bibinfo{person}{Jeremy Clark},
  \bibinfo{person}{Joshua~A. Kroll}, {and} \bibinfo{person}{Edward~W. Felten}.}
  \bibinfo{year}{2014}\natexlab{}.
\newblock \showarticletitle{Mixcoin: Anonymity for Bitcoin with accountable
  mixes}. In \bibinfo{booktitle}{\emph{Financial Cryptography 2014}}.
\newblock


\bibitem[\protect\citeauthoryear{Buterin et~al\mbox{.}}{Buterin
  et~al\mbox{.}}{2013}]%
        {buterin2013ethereum}
\bibfield{author}{\bibinfo{person}{Vitalik Buterin} {et~al\mbox{.}}}
  \bibinfo{year}{2013}\natexlab{}.
\newblock \bibinfo{title}{Ethereum white paper}.
\newblock
\newblock


\bibitem[\protect\citeauthoryear{Buterin and Reitwiessner}{Buterin and
  Reitwiessner}{2017}]%
        {eip197}
\bibfield{author}{\bibinfo{person}{Vitalik Buterin} {and}
  \bibinfo{person}{Christian Reitwiessner}.} \bibinfo{year}{2017}\natexlab{}.
\newblock \bibinfo{title}{Ethereum Improvement Proposal 197}.
\newblock
  \bibinfo{howpublished}{\url{https://github.com/ethereum/EIPs/blob/master/EIPS/eip-197.md}}.
\newblock


\bibitem[\protect\citeauthoryear{Castro, Liskov, et~al\mbox{.}}{Castro
  et~al\mbox{.}}{1999}]%
        {castro1999practical}
\bibfield{author}{\bibinfo{person}{Miguel Castro}, \bibinfo{person}{Barbara
  Liskov}, {et~al\mbox{.}}} \bibinfo{year}{1999}\natexlab{}.
\newblock \showarticletitle{Practical Byzantine fault tolerance}. In
  \bibinfo{booktitle}{\emph{OSDI}}, Vol.~\bibinfo{volume}{99}.
  \bibinfo{pages}{173--186}.
\newblock


\bibitem[\protect\citeauthoryear{Chaum}{Chaum}{1983}]%
        {chaum1983blind}
\bibfield{author}{\bibinfo{person}{David Chaum}.}
  \bibinfo{year}{1983}\natexlab{}.
\newblock \showarticletitle{Blind signatures for untraceable payments}. In
  \bibinfo{booktitle}{\emph{Advances in cryptology}}. Springer,
  \bibinfo{pages}{199--203}.
\newblock


\bibitem[\protect\citeauthoryear{Chen, Chen, Xiao, Zhang, Lin, and Lai}{Chen
  et~al\mbox{.}}{2019}]%
        {chen2019sgxpectre}
\bibfield{author}{\bibinfo{person}{Guoxing Chen}, \bibinfo{person}{Sanchuan
  Chen}, \bibinfo{person}{Yuan Xiao}, \bibinfo{person}{Yinqian Zhang},
  \bibinfo{person}{Zhiqiang Lin}, {and} \bibinfo{person}{Ten~H Lai}.}
  \bibinfo{year}{2019}\natexlab{}.
\newblock \showarticletitle{SgxPectre: Stealing Intel secrets from SGX enclaves
  via speculative execution}. In \bibinfo{booktitle}{\emph{2019 IEEE European
  Symposium on Security and Privacy (EuroS\&P)}}. IEEE,
  \bibinfo{pages}{142--157}.
\newblock


\bibitem[\protect\citeauthoryear{Community}{Community}{[n. d.]}]%
        {gas_station}
\bibfield{author}{\bibinfo{person}{Concourse~Open Community}.}
  \bibinfo{year}{[n. d.]}\natexlab{}.
\newblock \bibinfo{title}{{ETH Gas Station}.}
\newblock \bibinfo{howpublished}{\url{https://www.ethgasstation.info}}.
\newblock


\bibitem[\protect\citeauthoryear{Costan and Devadas}{Costan and
  Devadas}{2016}]%
        {costan2016intel}
\bibfield{author}{\bibinfo{person}{Victor Costan} {and}
  \bibinfo{person}{Srinivas Devadas}.} \bibinfo{year}{2016}\natexlab{}.
\newblock \showarticletitle{Intel SGX Explained.}
\newblock \bibinfo{journal}{\emph{IACR Cryptology ePrint Archive}}
  \bibinfo{volume}{2016}, \bibinfo{number}{086} (\bibinfo{year}{2016}),
  \bibinfo{pages}{1--118}.
\newblock


\bibitem[\protect\citeauthoryear{foundation}{foundation}{[n. d.]a}]%
        {web3py}
\bibfield{author}{\bibinfo{person}{Ethereum foundation}.} \bibinfo{year}{[n.
  d.]}\natexlab{a}.
\newblock \bibinfo{title}{A python interface for interacting with the Ethereum
  blockchain and ecosystem.}
\newblock \bibinfo{howpublished}{\url{https://github.com/ethereum/web3.py}}.
\newblock


\bibitem[\protect\citeauthoryear{foundation}{foundation}{[n. d.]b}]%
        {remix}
\bibfield{author}{\bibinfo{person}{Ethereum foundation}.} \bibinfo{year}{[n.
  d.]}\natexlab{b}.
\newblock \bibinfo{title}{Remix.}
\newblock \bibinfo{howpublished}{\url{https://remix.ethereum.org}}.
\newblock


\bibitem[\protect\citeauthoryear{Galal}{Galal}{[n. d.]}]%
        {auctioneer}
\bibfield{author}{\bibinfo{person}{Hisham~S. Galal}.} \bibinfo{year}{[n.
  d.]}\natexlab{}.
\newblock \bibinfo{title}{Auctioneer source code.}
\newblock
  \bibinfo{howpublished}{\url{https://github.com/HSG88/AuctionContract/tree/master/Auctioneer}}.
\newblock


\bibitem[\protect\citeauthoryear{Galal and Youssef}{Galal and Youssef}{2018a}]%
        {galal2018succinctly}
\bibfield{author}{\bibinfo{person}{Hisham~S Galal} {and} \bibinfo{person}{Amr~M
  Youssef}.} \bibinfo{year}{2018}\natexlab{a}.
\newblock \showarticletitle{Succinctly Verifiable Sealed-Bid Auction Smart
  Contract}.
\newblock In \bibinfo{booktitle}{\emph{Data Privacy Management,
  Cryptocurrencies and Blockchain Technology}}. \bibinfo{publisher}{Springer},
  \bibinfo{pages}{3--19}.
\newblock


\bibitem[\protect\citeauthoryear{Galal and Youssef}{Galal and Youssef}{2018b}]%
        {galal2018verifiable}
\bibfield{author}{\bibinfo{person}{Hisham~S Galal} {and} \bibinfo{person}{Amr~M
  Youssef}.} \bibinfo{year}{2018}\natexlab{b}.
\newblock \showarticletitle{Verifiable sealed-bid auction on the ethereum
  blockchain}. In \bibinfo{booktitle}{\emph{International Conference on
  Financial Cryptography and Data Security}}. Springer,
  \bibinfo{pages}{265--278}.
\newblock


\bibitem[\protect\citeauthoryear{Galbraith, Paterson, and Smart}{Galbraith
  et~al\mbox{.}}{2008}]%
        {galbraith2008pairings}
\bibfield{author}{\bibinfo{person}{Steven~D Galbraith},
  \bibinfo{person}{Kenneth~G Paterson}, {and} \bibinfo{person}{Nigel~P Smart}.}
  \bibinfo{year}{2008}\natexlab{}.
\newblock \showarticletitle{Pairings for cryptographers}.
\newblock \bibinfo{journal}{\emph{Discrete Applied Math.}}
  \bibinfo{volume}{156}, \bibinfo{number}{16} (\bibinfo{year}{2008}),
  \bibinfo{pages}{3113--3121}.
\newblock


\bibitem[\protect\citeauthoryear{Gennaro, Jarecki, Krawczyk, and Rabin}{Gennaro
  et~al\mbox{.}}{1999}]%
        {gennaro1999secure}
\bibfield{author}{\bibinfo{person}{Rosario Gennaro}, \bibinfo{person}{Stanislaw
  Jarecki}, \bibinfo{person}{Hugo Krawczyk}, {and} \bibinfo{person}{Tal
  Rabin}.} \bibinfo{year}{1999}\natexlab{}.
\newblock \showarticletitle{Secure distributed key generation for discrete-log
  based cryptosystems}. In \bibinfo{booktitle}{\emph{Eurocrypt}},
  Vol.~\bibinfo{volume}{99}. Springer, \bibinfo{pages}{295--310}.
\newblock


\bibitem[\protect\citeauthoryear{Harkavy, Tygar, and Kikuchi}{Harkavy
  et~al\mbox{.}}{1998}]%
        {harkavy1998electronic}
\bibfield{author}{\bibinfo{person}{Michael Harkavy}, \bibinfo{person}{J~Doug
  Tygar}, {and} \bibinfo{person}{Hiroaki Kikuchi}.}
  \bibinfo{year}{1998}\natexlab{}.
\newblock \showarticletitle{Electronic Auctions with Private Bids.}. In
  \bibinfo{booktitle}{\emph{USENIX Workshop on Electronic Commerce}}.
\newblock


\bibitem[\protect\citeauthoryear{Heilman, Alshenibr, Baldimtsi, Scafuro, and
  Goldberg}{Heilman et~al\mbox{.}}{2016}]%
        {tumblebit}
\bibfield{author}{\bibinfo{person}{Ethan Heilman}, \bibinfo{person}{Leen
  Alshenibr}, \bibinfo{person}{Foteini Baldimtsi}, \bibinfo{person}{Alessandra
  Scafuro}, {and} \bibinfo{person}{Sharon Goldberg}.}
  \bibinfo{year}{2016}\natexlab{}.
\newblock \showarticletitle{TumbleBit: An untrusted Bitcoin-compatible
  anonymous payment hub}. In \bibinfo{booktitle}{\emph{NDSS 2017}}.
\newblock


\bibitem[\protect\citeauthoryear{Kalodner, Goldfeder, Chen, Weinberg, and
  Felten}{Kalodner et~al\mbox{.}}{2018}]%
        {kalodner2018arbitrum}
\bibfield{author}{\bibinfo{person}{Harry Kalodner}, \bibinfo{person}{Steven
  Goldfeder}, \bibinfo{person}{Xiaoqi Chen}, \bibinfo{person}{S~Matthew
  Weinberg}, {and} \bibinfo{person}{Edward~W Felten}.}
  \bibinfo{year}{2018}\natexlab{}.
\newblock \showarticletitle{Arbitrum: Scalable, private smart contracts}. In
  \bibinfo{booktitle}{\emph{27th $\{$USENIX$\}$ Security Symposium
  ($\{$USENIX$\}$ Security 18)}}. \bibinfo{pages}{1353--1370}.
\newblock


\bibitem[\protect\citeauthoryear{Karame, Androulaki, and Capkun}{Karame
  et~al\mbox{.}}{2012}]%
        {karame2012double}
\bibfield{author}{\bibinfo{person}{Ghassan~O Karame}, \bibinfo{person}{Elli
  Androulaki}, {and} \bibinfo{person}{Srdjan Capkun}.}
  \bibinfo{year}{2012}\natexlab{}.
\newblock \showarticletitle{Double-spending fast payments in bitcoin}. In
  \bibinfo{booktitle}{\emph{ACM conference on Computer and communications
  security}}. ACM, \bibinfo{pages}{906--917}.
\newblock


\bibitem[\protect\citeauthoryear{Kate, Huang, and Goldberg}{Kate
  et~al\mbox{.}}{2012a}]%
        {kate2012distributed}
\bibfield{author}{\bibinfo{person}{Aniket Kate}, \bibinfo{person}{Yizhou
  Huang}, {and} \bibinfo{person}{Ian Goldberg}.}
  \bibinfo{year}{2012}\natexlab{a}.
\newblock \showarticletitle{Distributed Key Generation in the Wild.}
\newblock \bibinfo{journal}{\emph{IACR Cryptology ePrint Archive}}
  \bibinfo{volume}{2012} (\bibinfo{year}{2012}), \bibinfo{pages}{377}.
\newblock


\bibitem[\protect\citeauthoryear{Kate, Huang, and Goldberg}{Kate
  et~al\mbox{.}}{2012b}]%
        {cryptoeprint:2012:377}
\bibfield{author}{\bibinfo{person}{Aniket Kate}, \bibinfo{person}{Yizhou
  Huang}, {and} \bibinfo{person}{Ian Goldberg}.}
  \bibinfo{year}{2012}\natexlab{b}.
\newblock \bibinfo{title}{Distributed Key Generation in the Wild}.
\newblock \bibinfo{howpublished}{Cryptology ePrint Archive, Report 2012/377}.
\newblock
\newblock
\shownote{\url{https://eprint.iacr.org/2012/377}.}


\bibitem[\protect\citeauthoryear{Kosba, Miller, Shi, Wen, and
  Papamanthou}{Kosba et~al\mbox{.}}{2016}]%
        {kosba2016hawk}
\bibfield{author}{\bibinfo{person}{Ahmed Kosba}, \bibinfo{person}{Andrew
  Miller}, \bibinfo{person}{Elaine Shi}, \bibinfo{person}{Zikai Wen}, {and}
  \bibinfo{person}{Charalampos Papamanthou}.} \bibinfo{year}{2016}\natexlab{}.
\newblock \showarticletitle{Hawk: The blockchain model of cryptography and
  privacy-preserving smart contracts}. In \bibinfo{booktitle}{\emph{2016 IEEE
  symposium on security and privacy (SP)}}. IEEE, \bibinfo{pages}{839--858}.
\newblock


\bibitem[\protect\citeauthoryear{Król, Sonnino, and Argyrios}{Król
  et~al\mbox{.}}{2020}]%
        {pastrami}
\bibfield{author}{\bibinfo{person}{Michał Król}, \bibinfo{person}{Alberto
  Sonnino}, {and} \bibinfo{person}{Tasiopoulos Argyrios}.}
  \bibinfo{year}{2020}\natexlab{}.
\newblock \bibinfo{title}{Pastrami source code.}
\newblock
  \bibinfo{howpublished}{\url{https://github.com/harnen/FilecoinPricingMechanism}}.
\newblock


\bibitem[\protect\citeauthoryear{Labs}{Labs}{2018a}]%
        {filecoin}
\bibfield{author}{\bibinfo{person}{Protocol Labs}.}
  \bibinfo{year}{2018}\natexlab{a}.
\newblock \bibinfo{booktitle}{\emph{{FileCoin}: A Decentralized Storage
  Network}}.
\newblock \bibinfo{type}{{T}echnical {R}eport}.
\newblock


\bibitem[\protect\citeauthoryear{Labs}{Labs}{2018b}]%
        {storj}
\bibfield{author}{\bibinfo{person}{Storj Labs}.}
  \bibinfo{year}{2018}\natexlab{b}.
\newblock \bibinfo{title}{Storj: A Decentralized Cloud Storage
  NetworkFramework}.
\newblock \bibinfo{howpublished}{"\url{https://storj.io/storj.pdf}"}.
\newblock


\bibitem[\protect\citeauthoryear{Lavoie, Hendren, Desprez, and Correia}{Lavoie
  et~al\mbox{.}}{2019}]%
        {lavoie2018pando}
\bibfield{author}{\bibinfo{person}{Erick Lavoie}, \bibinfo{person}{Laurie~J.
  Hendren}, \bibinfo{person}{Frederic Desprez}, {and} \bibinfo{person}{Miguel
  Correia}.} \bibinfo{year}{2019}\natexlab{}.
\newblock \showarticletitle{{Pando}: Personal Volunteer Computing in Browsers}.
  In \bibinfo{booktitle}{\emph{20th {ACM/IFIP} International Conference on
  {Middleware}}}.
\newblock


\bibitem[\protect\citeauthoryear{Lehmann, M{\"u}ller, and Sandholm}{Lehmann
  et~al\mbox{.}}{2006}]%
        {lehmann2006winner}
\bibfield{author}{\bibinfo{person}{Daniel Lehmann}, \bibinfo{person}{Rudolf
  M{\"u}ller}, {and} \bibinfo{person}{Tuomas Sandholm}.}
  \bibinfo{year}{2006}\natexlab{}.
\newblock \showarticletitle{The winner determination problem}.
\newblock \bibinfo{journal}{\emph{Combinatorial auctions}}
  (\bibinfo{year}{2006}), \bibinfo{pages}{297--318}.
\newblock


\bibitem[\protect\citeauthoryear{Maxwell}{Maxwell}{2013}]%
        {coinjoin}
\bibfield{author}{\bibinfo{person}{Gregory Maxwell}.}
  \bibinfo{year}{2013}\natexlab{}.
\newblock \bibinfo{title}{CoinJoin: Bitcoin privacy for the real world}.
\newblock
  \bibinfo{howpublished}{\url{https://bitcointalk.org/index.php?topic=279249}}.
\newblock


\bibitem[\protect\citeauthoryear{Meiklejohn and Mercer}{Meiklejohn and
  Mercer}{2018}]%
        {mobius}
\bibfield{author}{\bibinfo{person}{Sarah Meiklejohn} {and}
  \bibinfo{person}{Rebekah Mercer}.} \bibinfo{year}{2018}\natexlab{}.
\newblock \showarticletitle{M{\"o}bius: Trustless tumbling for transaction
  privacy}.
\newblock \bibinfo{journal}{\emph{Proceedings on Privacy Enhancing
  Technologies}} \bibinfo{volume}{2018}, \bibinfo{number}{2}
  (\bibinfo{year}{2018}), \bibinfo{pages}{105--121}.
\newblock


\bibitem[\protect\citeauthoryear{Mishra and Parkes}{Mishra and Parkes}{2004}]%
        {mishra2004multi}
\bibfield{author}{\bibinfo{person}{Debasis Mishra} {and}
  \bibinfo{person}{David~C Parkes}.} \bibinfo{year}{2004}\natexlab{}.
\newblock \bibinfo{booktitle}{\emph{Multi-item {Vickrey-Dutch} auction for
  unit-demand preferences}}.
\newblock \bibinfo{type}{{T}echnical {R}eport}. \bibinfo{institution}{Technical
  Report Harvard EconCS Technical Report, Harvard University}.
\newblock


\bibitem[\protect\citeauthoryear{Mishra and Parkes}{Mishra and Parkes}{2009}]%
        {mishra2009multi}
\bibfield{author}{\bibinfo{person}{Debasis Mishra} {and}
  \bibinfo{person}{David~C Parkes}.} \bibinfo{year}{2009}\natexlab{}.
\newblock \showarticletitle{Multi-item {Vickrey-Dutch} auctions}.
\newblock \bibinfo{journal}{\emph{Games and Economic Behavior}}
  \bibinfo{volume}{66}, \bibinfo{number}{1} (\bibinfo{year}{2009}),
  \bibinfo{pages}{326--347}.
\newblock


\bibitem[\protect\citeauthoryear{Myerson and Satterthwaite}{Myerson and
  Satterthwaite}{1983}]%
        {myerson1983efficient}
\bibfield{author}{\bibinfo{person}{Roger~B Myerson} {and}
  \bibinfo{person}{Mark~A Satterthwaite}.} \bibinfo{year}{1983}\natexlab{}.
\newblock \showarticletitle{Efficient mechanisms for bilateral trading}.
\newblock \bibinfo{journal}{\emph{Journal of economic theory}}
  \bibinfo{volume}{29}, \bibinfo{number}{2} (\bibinfo{year}{1983}),
  \bibinfo{pages}{265--281}.
\newblock


\bibitem[\protect\citeauthoryear{Nakamoto}{Nakamoto}{2008}]%
        {bitcoin}
\bibfield{author}{\bibinfo{person}{Satoshi Nakamoto}.}
  \bibinfo{year}{2008}\natexlab{}.
\newblock \showarticletitle{Bitcoin: A peer-to-peer electronic cash system}.
\newblock  (\bibinfo{year}{2008}).
\newblock


\bibitem[\protect\citeauthoryear{project}{project}{[n. d.]}]%
        {filecoin_storage_issue}
\bibfield{author}{\bibinfo{person}{Filecoin project}.} \bibinfo{year}{[n.
  d.]}\natexlab{}.
\newblock \bibinfo{title}{Feature request: Clients can get estimates of miner
  storage.}
\newblock
  \bibinfo{howpublished}{\url{https://github.com/filecoin-project/go-filecoin/issues/2991}}.
\newblock


\bibitem[\protect\citeauthoryear{Ruffing, Moreno{-}Sanchez, and Kate}{Ruffing
  et~al\mbox{.}}{2014}]%
        {coinshuffle}
\bibfield{author}{\bibinfo{person}{Tim Ruffing}, \bibinfo{person}{Pedro
  Moreno{-}Sanchez}, {and} \bibinfo{person}{Aniket Kate}.}
  \bibinfo{year}{2014}\natexlab{}.
\newblock \showarticletitle{CoinShuffle: Practical Decentralized Coin Mixing
  for Bitcoin}. In \bibinfo{booktitle}{\emph{{ESORICS} {(2)}}}
  \emph{(\bibinfo{series}{Lecture Notes in Computer Science})},
  Vol.~\bibinfo{volume}{8713}. \bibinfo{publisher}{Springer},
  \bibinfo{pages}{345--364}.
\newblock


\bibitem[\protect\citeauthoryear{Shapley and Shubik}{Shapley and
  Shubik}{1971}]%
        {shapley1971assignment}
\bibfield{author}{\bibinfo{person}{Lloyd~S Shapley} {and}
  \bibinfo{person}{Martin Shubik}.} \bibinfo{year}{1971}\natexlab{}.
\newblock \showarticletitle{The assignment game I: The core}.
\newblock \bibinfo{journal}{\emph{International Journal of game theory}}
  \bibinfo{volume}{1}, \bibinfo{number}{1} (\bibinfo{year}{1971}),
  \bibinfo{pages}{111--130}.
\newblock


\bibitem[\protect\citeauthoryear{Sonnino, Al-Bassam, Bano, and Danezis}{Sonnino
  et~al\mbox{.}}{2018}]%
        {coconut}
\bibfield{author}{\bibinfo{person}{Alberto Sonnino}, \bibinfo{person}{Mustafa
  Al-Bassam}, \bibinfo{person}{Shehar Bano}, {and} \bibinfo{person}{George
  Danezis}.} \bibinfo{year}{2018}\natexlab{}.
\newblock \showarticletitle{Coconut: Threshold Issuance Selective Disclosure
  Credentials with Applications to Distributed Ledgers}.
\newblock \bibinfo{journal}{\emph{arXiv preprint arXiv:1802.07344}}
  (\bibinfo{year}{2018}).
\newblock


\bibitem[\protect\citeauthoryear{Storage}{Storage}{[n. d.]}]%
        {google_storage_cmp}
\bibfield{author}{\bibinfo{person}{Google~Cloud Storage}.} \bibinfo{year}{[n.
  d.]}\natexlab{}.
\newblock \bibinfo{title}{Pricing details by storage class.}
\newblock
  \bibinfo{howpublished}{\url{https://cloud.google.com/storage/pricing-summary/}}.
\newblock


\bibitem[\protect\citeauthoryear{Teutsch, Luu, and Reitwiessner}{Teutsch
  et~al\mbox{.}}{2016}]%
        {teutsch2016truebit}
\bibfield{author}{\bibinfo{person}{Jason Teutsch}, \bibinfo{person}{Loi Luu},
  {and} \bibinfo{person}{Christian Reitwiessner}.}
  \bibinfo{year}{2016}\natexlab{}.
\newblock \bibinfo{title}{Truebit: A verification and storage solution for
  blockchains}.
\newblock
\newblock


\bibitem[\protect\citeauthoryear{{The Golem Project}}{{The Golem
  Project}}{2016}]%
        {golem}
\bibfield{author}{\bibinfo{person}{{The Golem Project}}.}
  \bibinfo{year}{2016}\natexlab{}.
\newblock \bibinfo{title}{Golem Whitepaper}.
\newblock
  \bibinfo{howpublished}{\url{https://golem.network/doc/Golemwhitepaper.pdf}}.
\newblock


\bibitem[\protect\citeauthoryear{Valenta and Rowan}{Valenta and Rowan}{2015}]%
        {blindcoin}
\bibfield{author}{\bibinfo{person}{Luke Valenta} {and} \bibinfo{person}{Brendan
  Rowan}.} \bibinfo{year}{2015}\natexlab{}.
\newblock \showarticletitle{Blindcoin: Blinded, Accountable Mixes for Bitcoin}.
  In \bibinfo{booktitle}{\emph{Financial Cryptography and Data Security}},
  \bibfield{editor}{\bibinfo{person}{Michael Brenner}, \bibinfo{person}{Nicolas
  Christin}, \bibinfo{person}{Benjamin Johnson}, {and} \bibinfo{person}{Kurt
  Rohloff}} (Eds.). \bibinfo{pages}{112--126}.
\newblock
\showISBNx{978-3-662-48051-9}


\bibitem[\protect\citeauthoryear{Van~Bulck, Minkin, Weisse, Genkin, Kasikci,
  Piessens, Silberstein, Wenisch, Yarom, and Strackx}{Van~Bulck
  et~al\mbox{.}}{2018}]%
        {van2018foreshadow}
\bibfield{author}{\bibinfo{person}{Jo Van~Bulck}, \bibinfo{person}{Marina
  Minkin}, \bibinfo{person}{Ofir Weisse}, \bibinfo{person}{Daniel Genkin},
  \bibinfo{person}{Baris Kasikci}, \bibinfo{person}{Frank Piessens},
  \bibinfo{person}{Mark Silberstein}, \bibinfo{person}{Thomas~F Wenisch},
  \bibinfo{person}{Yuval Yarom}, {and} \bibinfo{person}{Raoul Strackx}.}
  \bibinfo{year}{2018}\natexlab{}.
\newblock \showarticletitle{Foreshadow: Extracting the keys to the intel {SGX}
  kingdom with transient out-of-order execution}. In
  \bibinfo{booktitle}{\emph{27th {USENIX} {Security} Symposium}}.
\newblock


\bibitem[\protect\citeauthoryear{Vickrey}{Vickrey}{1961}]%
        {vickrey1961counterspeculation}
\bibfield{author}{\bibinfo{person}{William Vickrey}.}
  \bibinfo{year}{1961}\natexlab{}.
\newblock \showarticletitle{Counterspeculation, auctions, and competitive
  sealed tenders}.
\newblock \bibinfo{journal}{\emph{The Journal of finance}}
  \bibinfo{volume}{16}, \bibinfo{number}{1} (\bibinfo{year}{1961}),
  \bibinfo{pages}{8--37}.
\newblock


\bibitem[\protect\citeauthoryear{Wang, Chen, Pan, Zhang, Wang, Bindschaedler,
  Tang, and Gunter}{Wang et~al\mbox{.}}{2017}]%
        {wang2017leaky}
\bibfield{author}{\bibinfo{person}{Wenhao Wang}, \bibinfo{person}{Guoxing
  Chen}, \bibinfo{person}{Xiaorui Pan}, \bibinfo{person}{Yinqian Zhang},
  \bibinfo{person}{XiaoFeng Wang}, \bibinfo{person}{Vincent Bindschaedler},
  \bibinfo{person}{Haixu Tang}, {and} \bibinfo{person}{Carl~A Gunter}.}
  \bibinfo{year}{2017}\natexlab{}.
\newblock \showarticletitle{Leaky cauldron on the dark land: Understanding
  memory side-channel hazards in SGX}. In \bibinfo{booktitle}{\emph{ACM SIGSAC
  Conference on Computer and Communications Security}}
  \emph{(\bibinfo{series}{CCS})}. ACM.
\newblock


\bibitem[\protect\citeauthoryear{Yuan, Xia, Chen, Zang, and Xie}{Yuan
  et~al\mbox{.}}{2018}]%
        {yuan2018shadoweth}
\bibfield{author}{\bibinfo{person}{Rui Yuan}, \bibinfo{person}{Yu-Bin Xia},
  \bibinfo{person}{Hai-Bo Chen}, \bibinfo{person}{Bin-Yu Zang}, {and}
  \bibinfo{person}{Jan Xie}.} \bibinfo{year}{2018}\natexlab{}.
\newblock \showarticletitle{ShadowEth: Private Smart Contract on Public
  Blockchain}.
\newblock \bibinfo{journal}{\emph{Journal of Computer Science and Technology}}
  \bibinfo{volume}{33}, \bibinfo{number}{3} (\bibinfo{year}{2018}),
  \bibinfo{pages}{542--556}.
\newblock


\end{thebibliography}
